\documentclass[prd,twocolumn,showpacs,floatfix,superscriptaddress,nofootinbib]{revtex4}
\usepackage[utf8]{inputenc}
\usepackage{graphicx}
\usepackage{epsfig}
\usepackage{bm}
\usepackage{amsfonts}
\usepackage[T1]{fontenc}
\usepackage{amssymb}
\usepackage{float}
\usepackage{amsmath}
\usepackage{dcolumn}
\usepackage{wasysym}
\usepackage{cancel}
\usepackage[colorlinks]{hyperref}
\usepackage[usenames,dvipsnames]{color}
\hypersetup{
     breaklinks=true,
    pdfstartview={FitH},    
    colorlinks=true,       
    linkcolor=blue,          
    citecolor=red,        
    filecolor=magenta,      
    urlcolor=blue,           
    anchorcolor=green,      
    linktocpage=true
}

\def\doi{http://doi.org}

\def\be{\begin{equation*}}
\def\ee{\end{equation*}}

\begin{document}

\title{Observational Tests of Regular Black Holes with Scalar Hair and their Stability}

\author{P. A. Gonz\'{a}lez}
\email{pablo.gonzalez@udp.cl} \affiliation{Facultad de
Ingenier\'{i}a y Ciencias, Universidad Diego Portales, Avenida Ej\'{e}rcito
Libertador 441, Casilla 298-V, Santiago, Chile.}
\author{Marco Olivares}
\email{marco.olivaresr@mail.udp.cl}
\affiliation{Facultad de
Ingenier\'{i}a y Ciencias, Universidad Diego Portales, Avenida Ej\'{e}rcito
Libertador 441, Casilla 298-V, Santiago, Chile.}
\author{Eleftherios Papantonopoulos}
\email{lpapa@central.ntua.gr}
\affiliation{Physics Division, School of Applied Mathematical and Physical Sciences, National Technical University of Athens, 15780 Zografou Campus, Athens, Greece.}
\author{Yerko V\'{a}squez}
\email{yvasquez@userena.cl}
\affiliation{Departamento de F\'{\i}sica, Facultad de Ciencias, Universidad de La Serena,\\
Avenida Cisternas 1200, La Serena, Chile.}

\begin{abstract}
    
We study the geodesic structure and observable properties of asymptotically flat regular black holes sourced by a phantom scalar field characterized by a scalar charge $A$. This parameter removes the central singularity and continuously deforms the Schwarzschild geometry. The equations of motion for test particles and photons are derived, and the resulting null geodesics are analyzed, including the deflection of light, gravitational time delay, and redshift, in order to constrain $A$ using classical Solar System tests. 
These observations impose stringent limits on the scalar charge, confirming that $A$ must remain extremely small in the weak-field regime to ensure full consistency with general relativity.
In the strong-field regime, we compute the Lyapunov exponent $\lambda$ associated with the photon sphere and establish its exact relations with the critical impact parameter $\mathcal{B}_u$ and the angular size of the shadow $\alpha_{\mathrm{sh}}$, given by $\mathcal{B}_u = 1/|\lambda|$ and $\alpha_{\mathrm{sh}} = 1/(r_{0}|\lambda|)$. 
These correspondences reveal that the dynamical instability of null circular orbits governs the optical appearance of the black hole. 
Our results show that increasing $A$ reduces the instability of photon trajectories and enlarges the angular size of the shadow, indicating that the regularization scale leaves a distinct observational imprint on the geometry of regular black holes. 
In addition, constraints derived from Event Horizon Telescope observations of M87* and Sgr~A* further restrict the allowed range of the scalar charge, reinforcing the consistency of the model with current astrophysical observations.

\end{abstract}

\maketitle

\tableofcontents

\section{Introduction}

In standard cosmology, after creation of matter at the end of inflation, the universe enters a deceleration regime and then, at late epochs, returns to an accelerated phase (the quintessence epoch) again. The recent observational results, on the other hand, indicate that in the early time cosmological evolution the formation of the matter structure was
governed by a peculiar matter, the dark energy, which was characterized by negative values of the pressure to density ratio $w$ which could even be $w < -1$ \cite{steinhardt,tegmark,seljak,hannestad,star03,chandra}. This kind of dark energy to give a negative value of $w$ should be parametrized by a {\it phantom field} having a negative kinetic energy \cite{sen,gorini,fara05}. However, in that case, a perfect-fluid description of dark energy is plagued with instabilities at small scales because of an imaginary velocity of sound that characterizes the phantom-matter case. To avoid this instability, a phantom scalar field may be regarded as an effective field description following from an underlying theory with positive energies \cite{no03,trod}.

Local solutions of a gravity theory with a minimally coupled scalar field with 
arbitrary potentials and negative kinetic energy were investigated in \cite{Bronnikov:2005gm}. It was found that
regular configurations were formed by the phantom scalar field in flat, de Sitter, and AdS asymptotic spacetimes,
avoiding the BH central singularity. Their main motivation was to find regular BH solutions with an expanding,
asymptotically de Sitter Kantowski-Sachs  cosmology beyond the event horizon. In the literature there are other solutions,
alternative to known ones, with a regular center~\cite{dym92,ned01,bdd03, Barrientos:2025rjn}.

Motivated by the above discussion in \cite{Karakasis:2023hni} of a gravity theory in the presence of a 
minimally coupled self-interacting phantom scalar field was studied. They looked for local solutions without gravitational singularities, which are points or regions of spacetime where a gravitational theory ceases to hold. Singularities are a common feature of BH physics. In most cases, the singularities are located at the center of the coordinate system, where curvature invariants possess a divergence and the spacetime is geodesically incomplete. However, Penrose showed that any singularity has to be covered by an event horizon (cosmic censorship hypothesis) and, as a result, all pathologies occurring at the singular region do not affect observers and physics outside of the horizon.

In \cite{Karakasis:2023hni}, in particular, a phantom scalar field with a scalar charge $A$ in a high-dimensional gravity theory was considered. If the scalar charge is zero, then the gravitational singularity is covered by a horizon, and then we have a normal BH with a constant scalar field. However, if $A$ is not zero, then the scalar charge of the phantom scalar field deforms the geometry in such a way that the gravitational singularity is absent and a compact object is generated with a horizon, which is a regular BH. A specific form of the metric function was used to derive the form of the self-interaction scalar potential by appropriately solving the system of the Lagrange equations of motion of the gravitational field theory. It was found that the  charge of the scalar field is connected to the mass of the BH dressing in this way the BH with secondary hair \footnote{We refer the reader to \cite{primary} for some examples of black hole solutions with a primary scalar hair.}. This results in the independence of the resulting scalar potentials on the compact object's mass, which in turn imposes restriction on the mass, and leads to its dependence on the scalar charge, thus rendering the hair secondary. It was also found that this is possible only for the
dimensionalities $D=3,4$ where the scalar potential depends only on the ratio of the scalar charge over the mass.

In a recent paper \cite{Farrah:2023opk} it was claimed that regular black holes (BHs) can be supported by astrophysical and cosmological observations as realistic astrophysical BH models, which can become cosmological at a large distance from the BH. In this way non-singular cosmological BH models can couple to the expansion of the universe, gaining mass proportional to the scale factor. This claim was based on a recent study of supermassive BHs within elliptical galaxies where it was found preferential BH growth, relative to galaxy stellar mass \cite{akiyama2022first}. This leads to a realistic behavior at infinity of BH models predicting that the gravitating mass of a BH can increase with the expansion of the universe independently of accretion or mergers, in a manner that depends on the BH interior solution. Then in \cite{Farrah:2023opk} it was proposed that stellar remnant BHs are the astrophysical origin of dark energy, which explained the onset of accelerating expansion of the universe.

Regular black holes supported by scalar fields constitute a well defined and extensively studied class of singularity-free solutions in general relativity. In particular, configurations sourced by phantom scalar fields provide analytically tractable geometries in which the central curvature singularity is removed while preserving asymptotic flatness. These models offer a controlled framework to investigate how regularization mechanisms modify the strong-field structure of black holes and their dynamical properties.

In the present work we focus on the class of asymptotically flat regular black holes constructed in Refs.~\cite{Bronnikov:2005gm, Karakasis:2023hni}, where a minimally coupled phantom scalar field generates a primary scalar charge that smooths out the central region of the spacetime. Our interest does not lie in cosmological coupling scenarios, but rather in exploring how the regularizing parameter affects geodesic motion. This setup provides a minimal and physically transparent laboratory to isolate the impact of scalar-induced regularization on black-hole dynamics. So, in particular, 
we consider a phantom scalar field with a scalar charge $A$ in a four-dimensional gravity theory \cite{Karakasis:2023hni}. As we discussed, if $A$ is not zero, then the scalar charge of the phantom scalar field deforms the geometry in such a way that the gravitational singularity is absent and a compact object is generated. We study the behavior of the null geodesics of an asymptotically flat regular black hole, see Refs.~\cite{Gonzalez:2015jna,Ramos:2021jta,Heydari-Fard:2021pjc,Theodosopoulos:2023ice,Chen:2024luw, Ovalle:2020kpd, Nekouee:2025zvp} for studies of geodesic motion in hairy black holes, Refs. \cite{Carvajal:2025ucx,Carvajal:2025emj,Gonzalez:2020vzl} in Horndeski black holes, and Refs.~\cite{Abbas:2014oua,Stuchlik:2014qja,Azam:2017adt,Azam:2017izk, Becerril:2020fek,Zhou:2022yio,Bautista-Olvera:2019blb,Xi:2023oib} for related analysis in regular black holes. Additionally,
to constrain the coupling parameter A of the scalar field to gravity, we consider three classical tests of gravity in the solar system, such as the bending of the light, the gravitational redshift, and the Shapiro time delay. Furthermore, through the Lyapunov exponents, we explore the stability of these geodesics in the presence of the coupling parameter A.

We find that larger values of A lead to a progressive stabilization of the photon sphere while the Schwarzschild case (A = 0) represents the most unstable configuration,  suggesting a connection between the regularity of the spacetime and its dynamical stability properties. Also, calculating the shadow of the black hole we find that the photon orbits become less unstable, indicating that the regularization parameter A not only smooths the central geometry but also stabilizes the photon sphere, thereby producing a more extended and less sharply defined shadow boundary.

The paper is organized as follows. 
In Sec.~\ref{sec:buildBH}, we present the setup of the theory describing regular black holes and discuss the main properties of the corresponding metric functions. 
Section~\ref{sec3} is devoted to the determination of the equations of motion for test particles and photons. 
In Sec.~\ref{NG}, we analyze the null geodesics for asymptotically flat regular black holes, including both the angular motion ($L\neq0$) and the purely radial case ($L=0$), and we explore several observational tests such as the deflection of light, the gravitational time delay, and the gravitational redshift. 
Section~\ref{SLE} introduces the Lyapunov exponents associated with the photon sphere and discusses their role as indicators of orbital stability. 
In Sec.~\ref{sec:connection_lyapunov_observables}, we establish the connection between the Lyapunov exponent and observational signatures, emphasizing its relation with the deflection of light and the black hole shadow. 
Finally, Sec.~\ref{FR} summarizes the main conclusions of this work.

\section{Setup of the theory of regular black holes}\label{sec:buildBH}

We consider the action
\begin{equation} S = \int d^4x \sqrt{-g}\left\{ \frac{R}{2\kappa} - \frac{1}{2}f(\phi)\nabla_{\mu}\phi\nabla^{\mu}\phi - V(\phi)\right\}~,\end{equation}
which consists of the Ricci scalar $R$~\footnote{Our conventions and definitions throughout this paper are: $(-,+,+,+)$ for the signature of the metric, the Riemann tensor is defined as
$R^\lambda_{\,\,\,\,\mu \nu \sigma} = \partial_\nu \, \Gamma^\lambda_{\,\,\mu\sigma} + \Gamma^\rho_{\,\, \mu\sigma} \, \Gamma^\lambda_{\,\, \rho\nu} - (\nu \leftrightarrow \sigma)$,
and the Ricci tensor and scalar are given by  $R_{\mu\nu} = R^\lambda_{\,\,\,\,\mu \lambda \nu}$ and $R= g^{\mu\nu}\, R_{\mu\nu}$ respectively.} and a self interacting scalar field, minimally coupled to gravity with $\kappa=8\pi G$ where we will set $G=1$ from now on. In the action, $f(\phi)$ is the function of the scalar field, being either a phantom $(f(\phi)<0)$ or a regular one $(f(\phi)>0)$. The field equations read
\begin{eqnarray}
&&G_{\mu\nu} =\kappa T_{\mu\nu} ~,\\
&&f(\phi)\Box\phi +\frac{f'(\phi)}{2}\nabla_{\mu}\phi\nabla^{\mu}\phi  = \frac{dV}{d\phi}~,\\
&&T_{\mu\nu} = f(\phi) \nabla_{\mu}\phi\nabla_{\nu}\phi - \frac{f(\phi)}{2}g_{\mu\nu}\nabla_{\alpha}\phi\nabla^{\alpha}\phi - V(\phi)~.
\end{eqnarray}
Considering the following metric ansatz
\begin{equation} ds^2 = - b(r)dt^2 + \cfrac{1}{b(r)}dr^2 + w(r)^2 d\Omega^2 \label{ds}~,\end{equation}
where
\begin{equation} d\Omega^{2} = d\theta^2 + \sin^{2}\theta d\varphi^2, \end{equation}
and
\begin{equation}
w(r) = \sqrt{r^2+A^2}~, \label{defor}
\end{equation}
where $A$ is a length scale, an ansatz that was first considered in \cite{Bronnikov:2005gm}.

In this section we consider the $D=4$ asymptotically flat regular black-hole solution constructed in Ref. \cite{Karakasis:2023hni}, which builds upon the phantom scalar configurations originally analyzed in Ref.  \cite{Bronnikov:2005gm}, where the corresponding energy conditions were also discussed, and  where all the possible solutions were presented. The stability of these solutions under radial perturbations was later studied in Ref. \cite{Bronnikov:2012ch}, where stable configurations were identified for particular values of the constants. Some properties of this solution like the gravitational lensing and the accretion process can be found in \cite{Ding:2013vta, Ditta2020}.

To solve the equations, we consider a phantom scalar field
\begin{align}\label{phant}
f(\phi)=-1\,,
\end{align}
and the solution in four dimensions reads
\begin{eqnarray}\label{bhor}
\notag b(r) &=& c_1 \left(A^2+r^2\right)\\
&&-\frac{c_2 \left(\left(A^2+r^2\right) \tan ^{-1}\left(\frac{r}{A}\right)+A r\right)+2 A r^2}{2 A^3}~,
\end{eqnarray}
and
\begin{eqnarray}\label{d4solution}
\phi(r) &=& \frac{1}{2 \sqrt{\pi }}\tan ^{-1}\left(\frac{r}{A}\right)~, \\
V(r) &=& \frac{c_2 \left(\left(A^2+3 r^2\right) \tan ^{-1}\left(\frac{r}{A}\right)+3 A r\right)}{16 \pi  A^3 \left(A^2+r^2\right)}\\
\notag && -\frac{2 A \left(A^2 c_1-1\right) \left(A^2+3 r^2\right)}{16 \pi  A^3 \left(A^2+r^2\right)}~,  \\
V(\phi) &=& \frac{4 A \left(A^2 c_1-1\right) \left(\cos \left(4 \sqrt{\pi } \phi \right)-2\right)}{32 \pi  A^3}\\
\notag && +\frac{c_2 \left(3 \sin \left(4 \sqrt{\pi } \phi \right)-4 \sqrt{\pi } \phi  \left(\cos \left(4 \sqrt{\pi } \phi \right)-2\right)\right)}{32 \pi  A^3}~.
\end{eqnarray}

The series expansion for the scalar field at infinity leads to
\begin{align}
\phi(r \to \infty) = \frac{\sqrt{\pi}}{4} - \frac{1}{2\sqrt{\pi}}\, \frac{A}{r} + \mathcal O\left(\frac{A^3}{r^3}\right)\, ,
\end{align}
from which we conclude that $A$ (or, to be precise, $-\frac{A}{2\sqrt{\pi}}$) controls the coefficient of the leading $1/r$ term in the scalar field and therefore plays the role of an asymptotic scalar charge associated with the solution.

The metric function $b(r)$ at infinity reads
\begin{eqnarray} 
\notag b(r\to\infty) &\sim& r^2 \left(c_1-\frac{4 A+\pi  c_2}{4 A^3}\right)+\left(A^2 c_1-\frac{\pi  c_2}{4 A}\right)\\
&&+\frac{c_2}{3 r}-\frac{A^2 c_2}{15
   r^3}+\mathcal{O}\left(\left(\frac{1}{r}\right)^5\right)~,
\end{eqnarray}
which resembles the (A)dS Schwarzschild black hole with corrections in the structure of spacetime that depend on the scalar charge $A$. Applying a transformation of the form $r^2 = R(r)^2 -A^2$, where $R$ will be the new coordinate, we find the following
\begin{eqnarray} 
\notag b(R\to\infty) &\sim& R^2 \left(-\frac{\pi  c_2}{4 A^3}-\frac{1}{A^2}+c_1\right)\\
&&+1+\frac{c_2}{3 R}+\mathcal{O}\left(\left(\frac{1}{R}\right)^3\right)\,.
\end{eqnarray}
It is clear that there is no deficit angle at large distances, and the solution describes a pure singularity-free black hole and not a gravitational monopole.
To make the spacetime asymptotically flat, we may set
\begin{equation} A^2 c_1-\frac{\pi  c_2}{4 A}=1 \to c_1=\frac{4 A+\pi  c_2}{4 A^3}~,\end{equation}
which fixes the value of $c_1$. Now, the asymptotic relation yields
\begin{equation} b(r) \sim 1+\frac{c_2}{3 r}-\frac{A^2 c_2}{15 r^3}+\frac{A^4 c_2}{35 r^5}+\mathcal{O}\left(\left(\frac{1}{r}\right)^7\right)~.\end{equation}
The metric above is clearly asymptotically flat and resembles the Schwarzschild black hole, while corrections in the structure of spacetime appear as $\mathcal{O}(r^{-n})$ terms (where $n \ge 3$) and are sourced by the phantom scalar charge $A$. The conserved black hole mass, can be calculated using the Abbott-Deser method, according to which, for an asymptotically flat spacetime, the mass $m$ is given by 
\begin{equation}\label{d4mass}
m = \frac{1}{2}\lim_{r\to\infty} r\left(\frac{1}{b(r)} -1\right) = -\frac{c_2}{6}~,
\end{equation}
and is clearly not affected by the scalar field. Then,  if one treats $m$ as independent of the scalar charge, the obtained spacetime describes a regular, asymptotically flat black hole, with a primary phantom scalar hair.

Now, setting $m=-c_2/6$, the potential reads
\begin{eqnarray} \label{d4pot}
\nonumber V(\phi) &=& \frac{3m}{16 \pi A^3} \Bigg[-8 \sqrt{\pi } \phi -3 \sin \left(4 \sqrt{\pi } \phi \right)\\
&&+2 \pi - \left(\pi -4 \sqrt{\pi } \phi \right) \cos \left(4 \sqrt{\pi } \phi \right)  \Bigg] ~.
\end{eqnarray}
while its asymptotic behavior reads
\begin{equation} V(r\to\infty) \sim \frac{A^2 m}{10 \pi  r^5}-\frac{13 \left(A^4 m\right)}{70 \pi  r^7}+\frac{9 A^6 m}{35 \pi  r^9}+\mathcal{O}\left(\left(\frac{1}{r}\right)^{11}\right)~.\end{equation}
Of course, for a negligible scalar charge, we obtain the Schwarzschild black hole.
\begin{equation}b(r,A\to0) \sim \left(1-\frac{2 m}{r}\right)+\frac{2 A^2 m}{5 r^3}+\mathcal{O}\left(A^4\right)~,
\end{equation}
and for vanishing mass $m$, we obtain a pure Minkowski spacetime, i.e. $b(r)=1$, hence $m$ is a genuine and independent scale of the solution. 
The horizon is obtained by solving $b(r)=0$, however, we cannot analytically solve this equation. Near the origin $b(r)$ behaves as
\begin{equation} b(r\to0)\sim \frac{4 A-6 \pi  m}{4 A}+\frac{6 m r}{A^2}-\frac{3 (\pi  m) r^2}{2 A^3}+\mathcal{O}\left(r^3\right)~,\end{equation}
where the dominant term is the constant term. 
For $b(r)$ to have a root in the region $0 \le r < \infty$, the constant term must be non-positive, since the term $\mathcal{O}(r)$ is non-negative. 
Hence, the scalar charge $A$ provides a bound for the mass $m$ of the black hole and the existence of a horizon
\begin{equation}\label{mAbound}
m \geq  \frac{2 A}{3 \pi }~.\end{equation}
In the $R$ coordinate, the line element reads
\begin{eqnarray}
\nonumber  ds^2 &=& -B(R)dt^2 +B(R)^{-1}\left(\frac{R^2-A^2}{R^2}\right)dR^2 + R^2d\Omega^2~, \\
\nonumber B(R) &=& 1+ \frac{3 m \sqrt{R^2-A^2}}{A^2}\\
&& -\frac{3 m R^2}{2 A^3} \left(\pi -2 \cot ^{-1}\left(\frac{A}{\sqrt{R^2-A^2}}\right)\right)~.
\end{eqnarray}
The stability of the system against radial perturbations was discussed \cite{Bronnikov:2012ch}, where it was found that the solutions are stable for particular values of the constants.\\

\begin{figure}[!h]
	\begin{center}
\includegraphics[width=80mm]{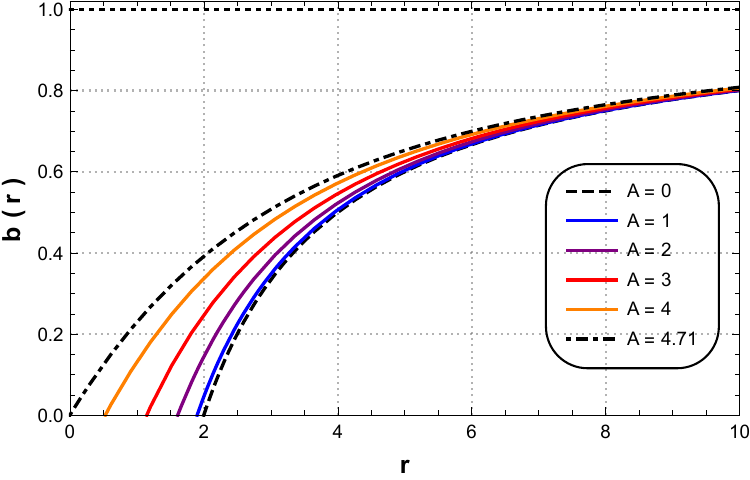}
	\end{center}
	\caption{Plot of the lapse function $b(r)$.
		Here we have used the value $m=1$. The event horizon is $r_+=2.000$ for $A=0$, $r_+=1.901$ for $A=1$, $r_+=1.610$ for $A=2$, $r_+=1.146$ for $A=3$, $r_+=0.526$ for $A=4$, and $r_+=0$ for $A=4.71$.}
	\label{f1}
\end{figure}

Asymptotically flat black hole. Now, we consider the asymptotically flat solution, where $c_1=\frac{1}{A^2}-\frac{3\pi m}{2A^3}$,  $c_2=-6m$ and $A \leq \frac{3\pi m}{2}$. The lapse function $b(r)$ is given by
\begin{eqnarray}\nonumber
b(r) &\equiv& \frac{3m}{A^2}r-\frac{r^2}{A^2}+\frac{(4A-6\pi m)(r^2+A^2)}{4\,A^{3}}+\\ \label{bflat0}
&+&\frac{3 m(r^2+A^2)}{\,A^{3}}\arctan{r\over A}\,.
\end{eqnarray}
If we consider the regime $A \ll r$, the lapse function $b(r)$ can be approximated by 
\begin{equation}
b(r) \approx  1-\frac{2m}{r}+\frac{2mA^2}{5\,r^{3}}\equiv \tilde{b}(r)\,,
\label{bflat1} 
\end{equation}
and the event horizon is obtained by solving $\tilde{b}(r) = 0$, which yields
\begin{equation}
r_+ =\frac{2m}{3}+\frac{4m}{3}\cos\left( {1\over 3}\arccos\left(1- {27A^2\over 40m^2}\right) \right) \,.
\label{horizon} 
\end{equation}
Note that when $A = 0$, the horizon reduces to the Schwarzschild black hole as expected.
Fig. \ref{f1} shows the radial profile of the lapse function $b(r)$ for different values of the parameter $A$. In the Schwarzschild limit ($A=0$), the lapse vanishes at the horizon, recovering the standard black hole solution. However, for $A>0$, the family of solutions corresponds to regular black holes, where the central singularity is removed while maintaining asymptotic flatness. Increasing values of $A$ shift the coordinate location of the event horizon toward smaller values of the radial coordinate $r$. However, since $r$ does not represent a physical distance in this chart, the invariant areal radius of the horizon, $\omega(r_+)$, actually increases with $A$, whereas the spacetime smoothly interpolates to the flat limit at large distances. At the critical value $A\simeq 4.71$, the lapse vanishes only at the limit, signaling a degenerate configuration that separates the domain of regular black holes with well-defined horizons from wormhole spacetimes.  


As we discussed in this section, the parameter A is a phantom scalar hair, and when the scalar field backreacts to the metric, A becomes a primary scalar charge. Above all, the most important role of this parameter is that it makes the black hole regular. In the next sections, we study how this parameter is constrained by observational results.

It is worth noting that the mass parameter $m$ appears explicitly in the scalar potential (\ref{d4pot}), which defines the theory. This implies that varying the black-hole mass modifies the underlying theory rather than simply selecting a different solution. An alternative formulation can be obtained through a straightforward reparametrization. Since the scalar potential depends only in the combination $m/A^3$, we introduce the constant
\[
c=\frac{m}{A^3},
\]
and regard $c$ as the parameter defining the theory. With this redefinition, the scalar potential becomes independent of the scalar charge $A$ and can be written as
\begin{eqnarray}
\notag V(\phi) &=& \frac{3 c}{16 \pi} \Bigg[-8 \sqrt{\pi } \phi -3 \sin \left(4 \sqrt{\pi } \phi \right)\\
&&+ 2 \pi - \left(\pi -4 \sqrt{\pi } \phi \right) \cos \left(4 \sqrt{\pi } \phi \right) \Bigg] \,,
\end{eqnarray}
while the metric function takes the form
\begin{eqnarray} 
\nonumber b(r) &=& 3 c A r-\frac{r^2}{A^2}+\frac{(4-6\pi c A^2)(r^2+A^2)}{4\,A^2}+\\
&+&3 c (r^2+A^2)\arctan{r\over A}\,.
\label{bflat0}
\end{eqnarray}
In this formulation, the theory is specified by the fixed constant $c$, and the mass is determined through $m=cA^3$. The scalar hair is therefore secondary, in the sense that it is not an independent parameter but is fixed once the black-hole solution is specified.

\section{Equations of motion}\label{sec3}

Having a regular black hole we will study how matter outside this regular black hole affects the background metric and in particular how the parameter A is constrained by the various observational effects of this matter.   

In order to obtain a description of the allowed motion of matter in the exterior spacetime of the black hole, we use the standard Lagrangian formalism \cite{Chandrasekhar:579245,Cruz:2004ts,Villanueva:2018kem}, so that the corresponding Lagrangian associated with the line element (\ref{ds}) reads

\begin{equation}\label{Lag1}
\mathcal{L}=-\frac{b(r)\, \dot{t}^2}{2}+ \frac{\dot{r}^2}{2\,b(r)}+\frac{\omega(r)^2}{2} \left(\dot{\theta}^2+ \sin^2 \theta\,\dot{\varphi} \right) \,,  \end{equation}
where $b(r)$ is the background metric (\ref{bhor}). Here, the dot indicates differentiation with respect to the affine parameter $\lambda$.  Since the Lagrangian (\ref{Lag1}) does not depend on the coordinates ($t,\varphi$), they are {\it cyclic coordinates} and, therefore, the corresponding conjugate momenta $\pi_{q} = \partial \mathcal{L}/\partial \dot{q}$ are conserved along the geodesic. Explicitly, we have 
\begin{eqnarray}\label{pp1}
\pi_{t}&=& -b(r)\, \dot{t}\equiv -E\,, \\\label{angcons} 
\pi_{\varphi}&=&\omega(r)^2\sin^2 \theta\, \dot{\varphi}\equiv L\,,
\end{eqnarray}
where $E$ denotes a positive constant associated with the time invariance of the Lagrangian, which can be identified with the conserved energy in the asymptotically flat limit of the spacetime defined by the line element (\ref{ds}). Similarly, 
$L$ corresponds to the conserved angular momentum, which implies that the motion is confined to an invariant plane. In this work, we restrict our analysis to the equatorial plane
$\theta= \pi/2$, such that $\dot\theta =0$.  Accordingly,  from Eq. (\ref{angcons}) we obtain.

\begin{equation}
 \dot{\varphi}=\frac{L}{\omega(r)^2}\,.
\label{eq1} 
\end{equation}
Therefore, using the fact that $2\mathcal{L}=-h^2$, where $h$ is the test mass,  together with Eqs. (\ref{pp1}) and (\ref{eq1}), we obtain the following equations of motion 

\begin{eqnarray}
\label{w.12}
&&\left(\frac{{\rm d}r}{{\rm d} \lambda}\right)^{2}= E^2-V^2(r)\,,\\
\label{w.13}
&&\left(\frac{{\rm d} r}{{\rm d} t}\right)^{2}= {b^{\,2}(r)\over E^2}\left[E^2-V^2(r)\right]\,,\\
\label{w.14}
&&\left(\frac{{\rm d} r}{{\rm d} \varphi}\right)^{2}=  {\omega(r)^{\,4}\over L^2}\left[E^2-V^2(r)\right]\,,
\end{eqnarray}
where the effective potential $V^2(r)$ is defined by
\begin{equation}
V^2(r) \equiv b(r)\left[h^2+\frac{L^{2}}{\omega(r)^{2}}\right]\,.
\label{Veff1} 
\end{equation}
Finally, by normalization, we shall consider $h=1$ for massive particles and $h=0$ for photons.

\section{Null geodesics}
\label{NG}
In order to obtain a description of the allowed motion for photons in the exterior spacetime of this regular black hole, we use the effective potential for asymptotically  flat  black holes,  in which the effective potential is obtained by setting $h=0$ in the above equation (\ref{Veff1}),
\begin{equation}\label{eqmot}
V^2(r)\equiv L^2\,\frac{b(r)}{\omega(r)^2}\,,
\end{equation}
\begin{figure}[!h]
	\begin{center}
		\includegraphics[width=80mm]{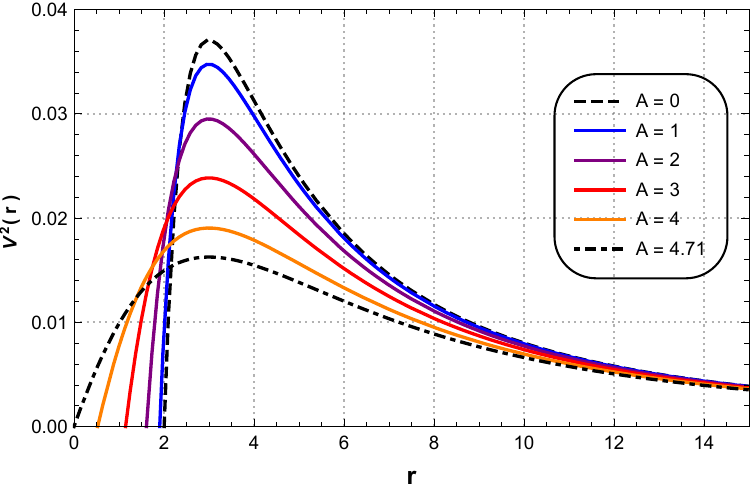}
	\end{center}
	\caption{Plot of the effective potential of photons.
		Here we have used the values $L=1$, $m=1$.}
	\label{f2}
\end{figure}
which is shown in Fig. \ref{f2}. The radius of the unstable circular orbit can be determined by the condition that the derivative of the effective potential vanishes. From Eq.~(\ref{eqmot}) we obtain
\begin{equation}\label{VeffLuz}
\dfrac{d V^2(r)}{dr} = - \frac{2L^2(r-3m)}{(A^2+r^2)^2}\,,
\end{equation}
it follows that the critical point occurs at $r_u=3m$,  which corresponds to the radius of the unstable circular photon orbit in $r$-chart.
Note that the radius of the unstable circular orbit remains unaffected by variations in the parameter $A$, and as
$A$ increases, the energy threshold required for a particle to reach such an orbit decreases, accompanied by a reduction in the radius of the event horizon, in this chart. However, the physical (areal) radius of the photon 
sphere is
\begin{equation}
R_u = \omega(3m) = \sqrt{9m^2 + A^2}\,.
\end{equation}
In particular, compared to Schwarzschild where $R_u^{\text{Sch}} = 3m$, 
the areal radius of the photon sphere in the present solution satisfies 
$R_u > 3m$ for any $A$, increasing monotonically with the scalar charge.

\begin{figure}[!h]
	\begin{center}
		\includegraphics[width=80mm]{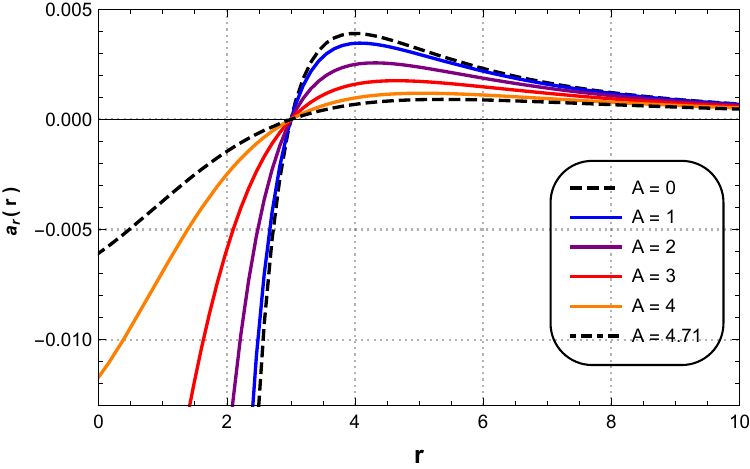}
	\end{center}
	\caption{Plot of the radial acceleration for massless particles. Here, $m=1$,  and $L=1$.
	The graph shows the radial acceleration. The radial acceleration is maximum at the inflection point of the effective potential. Note that for $r_+<r<r_u$, the  radial acceleration $a_{r}<0$, for $r=r_u$, the  radial acceleration $a_r=0$, for $r_u<r<\infty$, the  radial acceleration $a_r>0$.}
	\label{f2b}
\end{figure}

On the other hand, in Fig. \ref{f2b} we plot the radial acceleration for the massless particles given by $ a_{r} \equiv \ddot{r} =-V^{2\,\prime}(r)/2$. So, in order to calculate the value of $r$, where the radial acceleration is maximum ($r_I$), we consider $V^2\,''(r)$ given by 
\begin{equation}\label{Vpp}
\frac{d^2V^2(r)}{dr^2} =-2L^2{(A^2+3(4m-r)r)\over(A^2+ r^2)^3} \,,
\end{equation}
and the condition $V^{2\,''}(r) = 0 \quad\Rightarrow \quad  r^2-4mr-{A^2\over3}=0$, which corresponds to a quadratic equation whose positive solution is
\begin{equation}
r_{I}=2m\left( 1+\sqrt{1+\frac{A^{2}}{12m^{2}}}\right)\,.
\label{n8l}
\end{equation}
Fig. \ref{f2b} shows the radial acceleration $a_r(r)$ for different values of the parameter $A$. In the Schwarzschild limit ($A=0$), the acceleration vanishes at $r=3m$, corresponding to the photon sphere, and displays the well-known transition from a strong attractive regime at small radii ($r_+<r<r_u$) to a repulsive barrier before asymptotically approaching zero ($r_u<r<\infty$). For $A>0$, the position of the photon sphere remains fixed at $r=3m$, while the radii where the radial acceleration is maximum ($r_I$) increase and the repulsive peak decreases as $A$ increases. In addition, at the critical value $A\simeq 4.71$, the maximum repulsive
decreases. This confirms that, while the location of the photon sphere is universal, the strength of the radial acceleration is regulated by $A$, providing a clear signature of the regular character of these black hole solutions.
In the next analysis, we consider two kinds of motion, for $L=0$ (radial motion), and $L>0$ (angular motion) of the photons.

\subsection{Angular motion  ($L\neq0$)}\label{angmot}

Now we study the motion with $L\neq0$, 
so we put our attention in  Eq. (\ref{w.14}), which, after using (\ref{eqmot}), is conveniently written as
\begin{equation}\label{eqmotL}
\left(\frac{{\rm d}r}{{\rm d}\varphi}\right)^2=\frac{\omega^4}{\mathcal{B}^2}-\omega^2\,b(r) 
\,,
\end{equation}
where $\mathcal{B} \equiv L/E$ is the impact parameter. 
Next, based on the impact parameter values
and Fig. \ref{f2},
we present a brief qualitative description of the allowed
angular motions for photons in this regular black hole solution.

\begin{itemize}
	\item \emph{Capture zone}:
	If $0<\mathcal{B} <\mathcal{B}_{u}$, the photons fall inexorably
	to the  horizon $r_+$, or escape to infinity, depending on the initial conditions,
	and its cross section,
	$\sigma$, in these geometry is \cite{wald}
	\begin{eqnarray}
    \label{mr51a}
	\notag \sigma&=&\pi\,\mathcal{B}_u^2={ \pi\,(A^2+r_u^2)\over b(r_u)}\\
	&=&{2 \pi\,A^3\over 2A-3\pi m+6m\arctan{3m\over A}}.
	\end{eqnarray}
	Note that when $A = 0$, the cross section reduces to the Schwarzschild black hole $	\sigma=27\pi m^2$.
	\item \emph{Critical trajectories}:
	If $\mathcal{B}=\mathcal{B}_{u}$, photons can remain in unstable circular orbits of radius $r_{u}$.
	Therefore, photons that arrive from the initial distance$r_i$ ($r_+ < r_i< r_u$, or $r_u< r_i<\infty$)
can asymptotically fall to a circle of radius $r_{u}$.
	The affine period in this orbit is given by 
	\begin{equation}\label{p1}
	T_{\lambda}=\frac{2\pi\,(r_u^2+A^2)}{L}=\frac{2\pi\,(9m^2+A^2)}{L}\,,
	\end{equation}
	and the coordinate period is given by 
	\begin{equation}\label{p2}
	T_t=2\pi\,\mathcal{B}_u=2\pi \sqrt{r_u^2+A^2\over b(r_u)}\,.
	\end{equation}
	    \item \emph{Deflection zone.} If $\mathcal{B}_{u} < \mathcal{B} < \infty$ 
and $r_{d} \leq r < \infty$, orbits of the first kind are allowed. 
In this case, photons can originate from a finite distance or come from infinity
until they reach the distance $r = r_{d}$, which is the solution of the equation $V(r_{d}) = E$, 
and are then deflected.

	\item \emph{Second kind geodesic.} If $\mathcal{B}_u<\mathcal{B}<\infty$, and $r_+<r <r_f$, orbits of the second kind are allowed, with a return point in the range $r_+<r_f<r_u$, where $r = r_{f}$ is the solution of equation $V(r_{f}) = E$.
\end{itemize}

\subsubsection{The orbits}

In this section, we analyze the different kinds of orbit that test particles follow in regular compact objects with scalar hair for the following parameters
$m=1$ and $L=1$.

\begin{figure}[!h]
	\begin{center}
		\includegraphics[width=60mm]{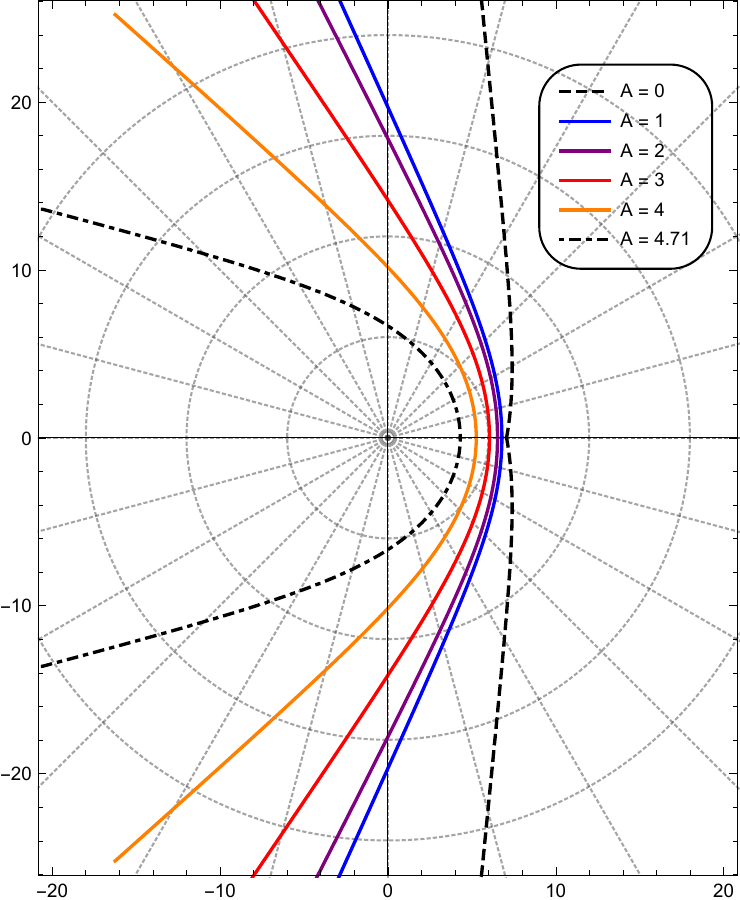}
	\end{center}
	\caption{Polar plot for deflection of light with $m=1$, and $L = 1$. All trajectories have the same energy $E^2=0.015$.}
	\label{f3}
\end{figure}

\begin{figure}[!h]
	\begin{center}
		\includegraphics[width=60mm]{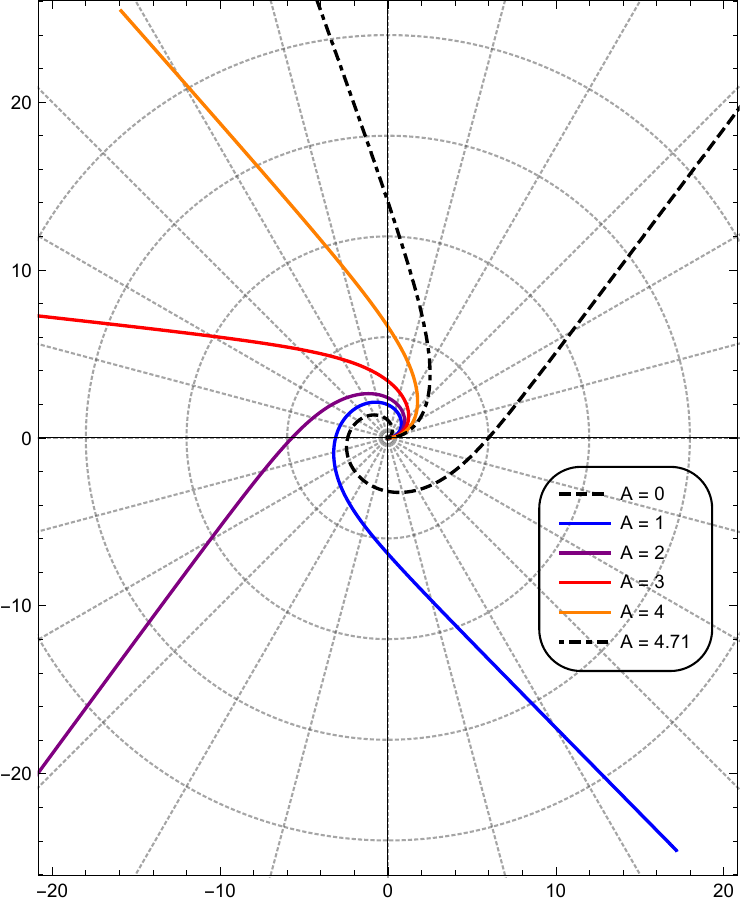}
	\end{center}
	\caption{The capture zone, trajectories  can plunge into the horizon or escape to infinity. Here, $m = 1$, $L = 1$, and $E^2=0.04$.}
	\label{captura}
\end{figure}

\paragraph{Deflection zone:} This zone presents orbits of the first kind, where the photons can come from a finite distance or from an infinity distance until they reach the distance $r = r_d$, and then the photons are deflected. Fig.~\ref{f3} shows the polar trajectories of light rays for different values of the parameter $A$, with fixed $m=1$, $L=1$, and the same energy $E^2=0.015$. This energy value allows the deflection of the light for all the values of $A$ considered. The Schwarzschild case ($A=0$) reproduces the standard deflection of light (dashed curve). As $A$ increases, the bending angle increases, with a return point that decreases as $A$ increases. This shows that the parameter $A$ directly controls the strength of the light deflection while preserving the asymptotic structure of spacetime.
\paragraph{Capture zone}
Photons with an impact parameter smaller than the critical one ($\mathcal{B} <\mathcal{B}_{u}$), which are in the capture zone, can plunge into the horizon or escape to infinity. So, by manipulating Eq. (\ref{eqmotL}), it allows us to show the trajectories of the photons in the capture zone in Fig. \ref{captura},
which illustrates their behavior for different values of the parameter $A$. 
All trajectories shown correspond to particles with the same energy ($E^{2}=0.04$), but their dynamics are strongly influenced by the effective potential barrier. 
As discussed in Fig.~\ref{f2}, the height of the potential maximum decreases with increasing $A$, which explains the differences in the trajectories. 
For $A=0$ (Schwarzschild), capture occurs after a slight deflection, whereas for the limiting case $A=4.71$ reduction of the potential barrier leads to a more direct plunge. 
Therefore, Fig.~\ref{captura} shows that increasing $A$ does not prevent capture, but changes the way it happens, evolving from moderately deflected paths to a more direct plunge.
\bigskip

\begin{figure}[!h]
	\begin{center}
		\includegraphics[width=70mm]{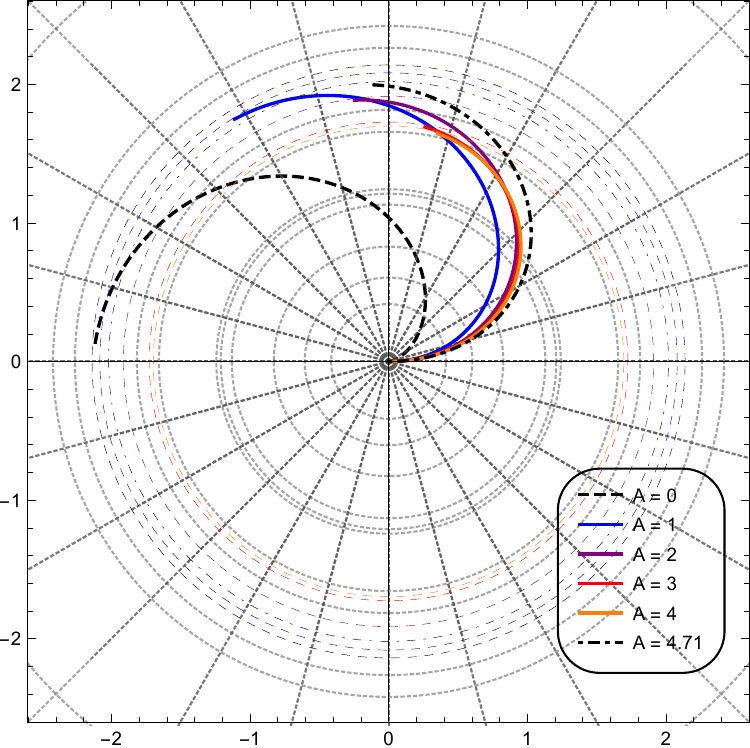}
	\end{center}
	\caption{Polar plot for second kind trajectories with $m=1$,  and $L = 1$. All trajectories have the same energy $E^2=0.015$.}
	\label{caida}
\end{figure}

\paragraph{Second kind trajectories}
The spacetime allows second kind trajectories, where the return point lies in the range $r_+ <r <r_0$, after which photons inevitably plunge into the horizon. In Fig. \ref{caida} we show the behavior of the second kind trajectories for a fixed value of the black hole mass $m$, the angular momentum $L$, and the energy $E$.
For $A=0$ (Schwarzschild), the trajectories approach the horizon relatively quickly, showing a short spiral. 
However, as $A$ increases, the radius of the horizon decreases (see Fig.~\ref{f1}) and the effective potential barrier is reduced (Fig.~\ref{f2}), which shifts the turning point and lengthens the spiral-like behavior. 
Consequently, photons with larger $A$ values undergo more revolutions around the regular black hole before being captured.
In the limiting case $A=4.71$, the path develops into a long spiral with multiple windings before the final plunge. 
Therefore, Fig.~\ref{caida} illustrates that the parameter $A$ governs not only the location of the horizon but also the degree of spiraling in second kind trajectories, enhancing the number of revolutions before the capture. 

\bigskip

\paragraph{Critical trajectories}

\begin{figure}[!h]
	\begin{center}
        \includegraphics[width=40mm]{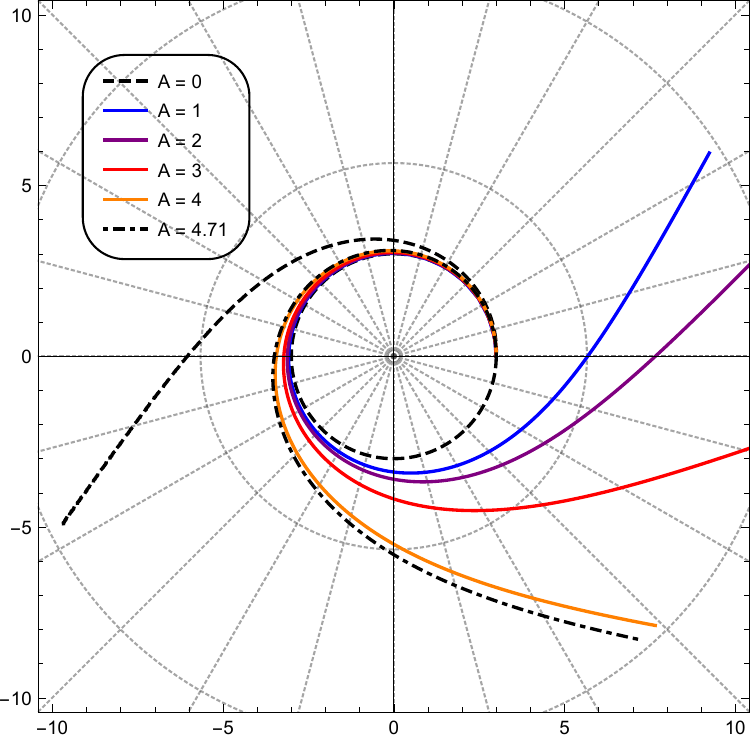}
        \includegraphics[width=40mm]{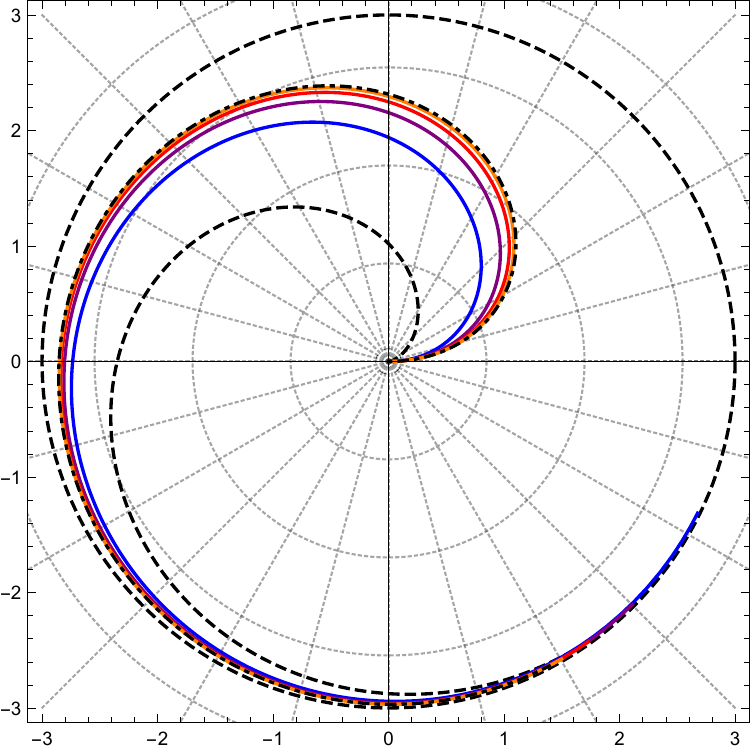}
	\end{center}
	\caption{The critical trajectories  plotted for $m = 1$  and $L = 1$. Left panel for CFK and right panel for CSK trajectories. Here, $E^2_u(A=0)=0.0373; E^2_u(A=1)= 0.0347; E^2_u(A=2)= 0.0295; E^2_u(A=3)= 0.0238; E^2_u(A=4)= 0.0190; E^2_u(A=4.71)= 0.0163$.}
	\label{criticas}
\end{figure}

In the case of $\mathcal{B}=\mathcal{B}_{u}$, the photons can be confined to unstable circular orbits of radius $r_u=3m$. This kind of motion is indeed ramified into two cases; critical trajectories of the first kind (CFK) in which the particles come from infinity to $r_u$ and those of the second kind (CSK) in which the photons approach at $r_u$ from distances smaller than $r_u$. 
In Fig. \ref{criticas}, we show the behavior of the CFK and CSK trajectories. The plots reveal that the effect of the parameter $A$ is to extend the spiraling behavior around the unstable orbit. 
For $A=0$ (Schwarzschild), both the CFK and CSK approach $r_{u}$ relatively quickly, developing short spirals. 
However, as $A$ increases, the reduction of the effective potential barrier (Fig.~\ref{f2}) and the inward shift of the horizon (Fig.~\ref{f1}) enhance the number of windings around $r_{u}$. 
In particular, the right panel of Fig.~\ref{criticas} shows that the CSK trajectories develop longer spirals for larger $A$, while the left panel demonstrates that CFK trajectories also take more revolutions before being captured. 
Therefore, the parameter $A$ controls the critical behavior of the photons, increasing the time spent near the unstable orbit and enlarging the spiraling structure of both the CFK and CSK trajectories.

\bigskip

\subsection{Motion with $L=0$}

\begin{figure}[!h]
	\begin{center}
		\includegraphics[width=80mm]{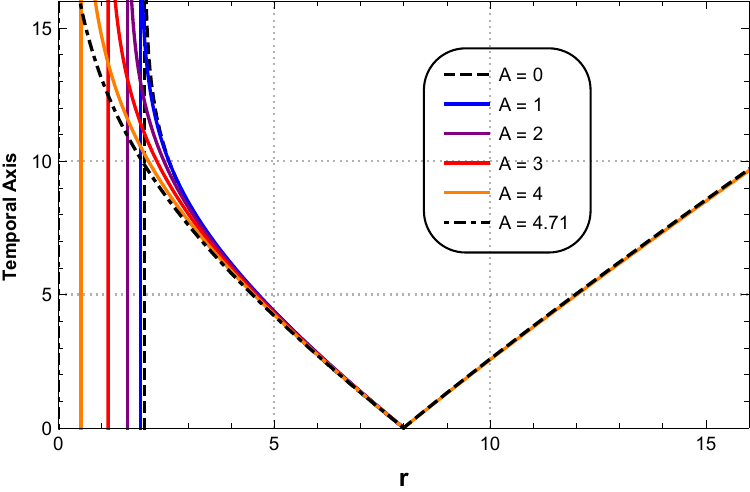}
	\end{center}
	\caption{Variation of the coordinate time $t$ as a function of the radial coordinate $r$ with $L=0$, $m=1$, $\rho_i = 8$, and different values of the regularization parameter $A = 0, 1, 2, 3, 4, 4.71$. 
The curves describe the coordinate time experienced by a photon as a test particle either falling towards the black hole or moving outward to infinity. 
Vertical dashed lines indicate the corresponding locations of the event horizons for each value of $A$.}
	\label{f444}
\end{figure}

For the motion of photons with null angular momentum, they are destined
to fall towards the event horizon or escape to infinity. From Eq.
(\ref{eqmot}) we have
$V^2(r)=0$. So, 
choosing the initial conditions for the photons as $r=\rho_i$
when $t=\lambda=0$, Eq. (\ref{w.12}) yields
\begin{equation}
\lambda(r)=\pm \frac{1}{E}(r-\rho_i)~,
\label{mr.3}
\end{equation}
where the ($-$) sign corresponds to photons falling into the event horizon and the ($+$) sign corresponds to photons that escape to infinity, and note that the above equation depends on the energy $E$. Also, for the negative sign the equation gives the result that the photons arrive at the event horizon in a finite affine parameter $\lambda_+=(\rho_i-r_+)/E$. On the other hand, 
from Eq. (\ref{w.13}) we obtain
\begin{equation}
t(r)=\pm \int^{r}_{\rho_i}{dr\over b(r)}\,,
\label{mr.5}
\end{equation}
which we compute numerically. The behavior of the coordinate time $t(r)$ for the motion of photons with $L=0$ is shown for different values of the parameter $A = 0, 1, 2, 3, 4, 4.71$, with $m=1$ and $\rho_i = 8$, in Fig. \ref{f444}. 
For $A = 0$, corresponding to the Schwarzschild limit, the coordinate time $t(r)$ 
diverges as the photon approaches the event horizon, reproducing the well-known behavior
perceived by a distant observer. As the parameter $A$ increases, the radius of the event horizon
$r_{+}$ decreases in $r$-chart, the invariant areal radius of the horizon $\omega(r_+)$ increases and the coordinate time still diverges at the horizon, which means that a photon
takes an infinite coordinate time to reach it or escape to spatial infinity. Consequently, a distant
observer never sees the photon crossing the horizon. Therefore, the causal structure at large distances
remains qualitatively similar to that of the Schwarzschild spacetime, while the parameter $A$ mainly
modifies the near-horizon geometry by shifting $r_{+}$ to smaller values. It should be noted that the
coordinate time $t$ is independent of the photon energy.

\subsection{Observational test}
\subsubsection{Deflection of light}
The deflection of light is important because the deflection of light by the Sun is one of the most important tests of general relativity, and the deflection of light by galaxies is the mechanism behind gravitational lenses. 
The effective potential $V^2(r)$ is 
\begin{equation}\label{VeffLuz4}
V^2(r)\approx \frac{L^2}{r^2+A^2}  \left( 1-\frac{2m}{r}+\frac{2mA^2}{5\,r^{3}}+\mathcal{O}(A^4)\right) \,.
\end{equation}

In this section, we will follow the procedure established in Ref. \cite{Straumann}. So, Eq. (\ref{w.14}) for photons is

\begin{eqnarray}\nonumber
\left(\frac{{\rm d}r}{{\rm d}\varphi}\right)^2&=&\frac{r^4}{\mathcal{B}^2}+\left(\frac{2A^2}{\mathcal{B}^2}-1\right)\,r^2+2m\,r+\\ \label{eqmot1}
&-&A^2+\frac{8mA^2}{r}+\mathcal{O}(A^4)\,,
\end{eqnarray}

Using the change of variables $r=1/u$, the above equation can be written as
\begin{equation}\label{du}
\left(\frac{du}{d\varphi}\right)^2=\frac{1}{\mathcal{B}^2}- \left(1-\frac{2A^2}{\mathcal{B}^2} \right) u^2+2m\, u^3-A^2u^4+\mathcal{O}(u^5)\,.
\end{equation}
Notice that for $A=0$, the above equation reduces to the classical equation of Schwarzschild for the motion of photons.
So, the derivative of Eq. (\ref{du}) with respect to $\varphi$ yields 
\begin{equation}\label{Binet1}
u^{\prime\prime}+ u =\frac{2A^2}{\mathcal{B}^2}u+3mu^2-2A^2u^3\,,
\end{equation}
where ${}^\prime$ denotes the derivative with respect to $\varphi$. By introducing the small dimensionless quantity $\epsilon=3m$, Eq. (\ref{Binet1}) can be rewritten as

\begin{equation}
\label{Binet2}
u''+u=\frac{2A^2}{3m\mathcal{B}^2}\epsilon \,u+\epsilon \,u^2-{2A^2\over 3m}\epsilon\,u^3\,.
\end{equation}
This can be solved by assuming the solution ansatz $u(\varphi)\approx u_0(\varphi)+\epsilon v(\varphi)+\mathcal{O}(\epsilon^2)$. So, substituting in the differential equation (\ref{Binet2}), we obtain
\begin{equation}
\label{Binet3}
u_0''+u_0+\epsilon(v''+v)=\frac{2A^2}{3m\mathcal{B}^2}\epsilon \,u_0+\epsilon \,u_0^2-{2A^2\over 3m}\epsilon\,u_0^3+\mathcal{O}(\epsilon^2)\,.
\end{equation}
Equating the zeroth-order terms in $\epsilon$, yields
\begin{equation}
u_0''+u_0=0\,.
\end{equation}
The solution for $u_0$ is 
\begin{equation}
\label{B0}
u_0=\frac{1}{\mathcal{B}}\sin\varphi\,.
\end{equation}
Now, equating the first-order $\epsilon$ terms in Eq. (\ref{Binet3}), yields

\begin{eqnarray}
v''+v&=&\frac{2A^2}{3m\mathcal{B}^2} \,u_0+\,u_0^2-{2A^2\over 3m}\,u_0^3\\ \label{v1}
\nonumber &=&\frac{2A^2}{3m\mathcal{B}^3} \,\sin\varphi+\,{(\sin\varphi)^2 \over \mathcal{B}^2 }-{2A^2\over 3m \mathcal{B}^3}\,(\sin\varphi)^3\,,
\end{eqnarray}
Thus, substituting $v=\bar{A}+\bar{B}\cos 2\varphi $ into Eq. (\ref{v1}), and using the trigonometric identity $2\sin^2\varphi=1-\cos2\varphi$, we obtain 
\begin{eqnarray}\nonumber
\bar{A}-3\bar{B}\cos \notag 2\varphi&=&\frac{\sqrt{2}\,A^2}{3m\mathcal{B}^3} \,(1-\cos 2\varphi)^{1/2}+\,{1\over 2\mathcal{B}^2 }(1-\cos 2\varphi)\\
\notag &&-{\sqrt{2}A^2\over 6m \mathcal{B}^3}\,(1-\cos 2\varphi)^{3/2}\\\label{m0}
\notag &\approx&\frac{\sqrt{2}\,A^2}{3m\mathcal{B}^3} \,(1-{1\over 2}\cos 2\varphi)+\,{1\over 2\mathcal{B}^2 }(1-\cos 2\varphi)\\
&&-{\sqrt{2}\,A^2\over 6m \mathcal{B}^3}\,(1-{3\over 2}\cos 2\varphi)\,,
\end{eqnarray}
where
\begin{equation}
\bar{A}=\frac{1}{2\mathcal{B}^2}+\frac{\sqrt{2}\,A^2}{6m\mathcal{B}^3}\,,
\end{equation}
\begin{equation}
\bar{B}=\frac{1}{6\mathcal{B}^2}-\frac{\sqrt{2}\,A^2}{36m\mathcal{B}^3}\,.
\end{equation}

The solution to Eq. (\ref{Binet1}) can be written as $u=u_0+3m(\bar{A}+\bar{B}\cos 2\varphi )$. Now, neglecting the last term, we obtain

\begin{equation}
u=\frac{1}{\mathcal{B}}\sin(\varphi)+\frac{3m}{2\mathcal{B}^2}+\frac{\sqrt{2}\,A^2}{2\mathcal{B}^3}
+\left(\frac{m}{2\mathcal{B}^2}-\frac{\sqrt{2}\,A^2}{12\mathcal{B}^3}\right)\cos(2\varphi)\,, 
\end{equation}
For large $r$ (small $u$), $\varphi$ is small, and we may take $\sin(\varphi)\approx \varphi$ and $\cos(\varphi)\approx 1$. In the limit $u \rightarrow 0$, $\varphi$ approaches $\varphi_{\infty}$, with
\begin{equation}
-\varphi_{\infty}=\frac{2m}{\mathcal{B}}+\frac{5\sqrt{2}\,A^2}{12\,\mathcal{B}^2}\,.
\end{equation}
Therefore, for the regular black hole considered, the deflection of light $\hat{\alpha}= 2\left |-\varphi_{\infty}\right |$ yields 
\begin{equation}\label{GB1}
\hat{\alpha}=\frac{4m}{\mathcal{B}}+\frac{5\sqrt{2}\,A^2}{6\,\mathcal{B}^2}\,.
\end{equation} 
Notice that $A=0$ recovers the classical result of GR $\hat{\alpha}=4m/\mathcal{B}$. 
Assuming the Sun as the central massive object, the deflection angle predicted by the Schwarzschild solution is given by
$\hat{\alpha}_{\mathrm{Sch}} = \frac{4M_{\odot}}{R_{\odot}}$
which, in arcsecond, reads
$
\hat{\delta}_{\mathrm{Sch}} = \frac{3600 \cdot 180}{\pi} \hat{\alpha}_{\mathrm{Sch}}=1.75092 \, \text{arcsec}\,. 
$
Observations near the Sun report measured values of
$\hat{\alpha}_P = 1.7520\, \text{arcsec}$ (prograde) and $\hat{\alpha}_R = 1.7519\, \text{arcsec}$ (retrograde)~\cite{Roy:2019ijp,Fathi:2025byw}. Therefore, by identifying the theoretical prediction with the observational value, we obtain the following constraints for $A$:
\begin{align}
\hat{\delta}_{\mathrm{RBH}} = \hat{\alpha}_P &\quad \Rightarrow \quad A = 46341\, \text{m}\,, \\
\hat{\delta}_{\mathrm{RBH}} = \hat{\alpha}_R &\quad \Rightarrow \quad A = 44139\, \text{m}\,,
\end{align}
where $\hat{\delta}_{\mathrm{RBH}} = \frac{3600 \cdot 180}{\pi} \hat{\alpha}$.

\subsubsection{Gravitational time delay}

An interesting relativistic phenomenon affecting the propagation of light rays is the apparent delay in the travel time of a light signal passing near the Sun, known as the Shapiro time delay.
This effect constitutes an important correction in high-precision astronomical observations.
The so-called time delay of radar echoes refers to the determination of the additional time taken by radar signals transmitted from Earth, passing through a region close to the Sun, to reach another planet or spacecraft and then return after reflection.
The total time interval between the emission and the reception of the pulse, as measured by a clock on Earth is

\begin{equation}
t_{12}=2\, t(r_1,\rho_0)+2\, t(r_2,\rho_0),
\end{equation}
where the signal is transmitted from a point 1 with coordinate $r_1$ to another point 2 with coordinate $r_2$, and $\rho_0$ as closest approach to the Sun. Now, in order to calculate the time delay we use (\ref{w.12}) and the coordinate time
\begin{equation}\label{ct}
\left(\frac{{\rm d} r}{{\rm d} t}\right)^{2}= b^{\,2}(r)\left[1-{V^2(r)\over E^2}\right]  \,.
\end{equation}  
At distance $r=\rho_0$, $dr/dt$ vanishes, so that
\begin{equation}\label{TD1}
\frac{E^2}{L^2}=\frac{b(\rho_0)}{\omega(\rho_0)^2}  \,.
\end{equation}
Now, by inserting (\ref{TD1}) in (\ref{ct}), the coordinate time that light requires to go from $\rho_0$ to $r$ is
\begin{equation}
t(r,\rho_0)=\int_{\rho_0}^r \frac{dr}{b(r)\sqrt{1-\frac{\omega(\rho_0)^2}{b(\rho_0)}\frac{b(r)}{\omega(r)^2}}}  \,.
\end{equation}
So, at first-order correction we obtain
\begin{eqnarray}
\nonumber t(r, \rho_0)&=&\sqrt{r^2-\rho_0^2}+m\sqrt{\frac{r-\rho_0}{r+\rho_0}}+2m\ln\left(\frac{r+\sqrt{r^2-\rho_0^2}}{\rho_0}\right)\\
\notag && +\frac{A^2}{2\rho_0}\sec^{-1}\left( {r\over \rho_0}\right) 
-\frac{mA^2}{2\rho_0^2}\sec^{-1}\left( {r\over \rho_0}\right) \\
&& +\frac{13mA^2}{10\rho_0^2} \sqrt{\frac{r-\rho_0}{r+\rho_0}}+\frac{9mA^2}{10\rho_0^2}{\sqrt{r^2-\rho_0^2}\over r}\,.
\end{eqnarray}
Therefore, for the circuit from point 1 ($r_1$) to point 2 ($r_2$) and back, the delay in coordinate time is
\begin{eqnarray}
\notag \Delta t&:=&  2\left[t(r_1, \rho_0)+t(r_2,\rho_0)-\sqrt{r_1^2-\rho_0^2}-\sqrt{r_2^2-\rho_0^2}\right]\\
&=&\Delta t_m+\Delta t_A+\Delta t_{mA},
\end{eqnarray}
where
\begin{eqnarray}
\notag \Delta t_m&=&2m\Big[ 2\, ln\left(\frac{(r_1+\sqrt{r_1^2-\rho_0^2})(r_2+\sqrt{r_2^2-\rho_0^2})}{\rho_0^2}\right)\\
&& +\sqrt{\frac{r_1-\rho_0}{r_1+\rho_0}}+\sqrt{\frac{r_2-\rho_0}{r_2+\rho_0}}\Big]\,,\\
\Delta t_A&=&\frac{A^2}{\rho_0}\left( \sec^{-1}\left( {r_1\over \rho_0}\right) +\sec^{-1}\left( {r_2\over \rho_0}\right)\right)\,,\\
\notag \Delta t_{mA}&=&\frac{13mA^2}{5\rho_0^2}\left[ \sqrt{\frac{r_1-\rho_0}{r_1+\rho_0}}+\sqrt{\frac{r_2-\rho_0}{r_2+\rho_0}}\right]\\
\notag && +\frac{9mA^2}{5\rho_0^2}\left( {\sqrt{r_1^2-\rho_0^2}\over r_1}+{\sqrt{r_2^2-\rho_0^2}\over r_2}\right) +\\
&-&\frac{mA^2}{\rho_0^2}\left( \sec^{-1}\left( {r_1\over \rho_0}\right) +\sec^{-1}\left( {r_2\over \rho_0}\right)\right)\,. 
\end{eqnarray}
Now, considering 
$\rho_0<<r_1,r_2$, we obtain  
\begin{eqnarray}
\notag \Delta t &\approx& 4m\left[ 1+ ln\left(\frac{4r_1r_2}{\rho_0^2}\right)\right]\\
&& +\frac{A^2}{\rho_0}\left( \sec^{-1}\left( {r_1\over \rho_0}\right) +\sec^{-1}\left( {r_2\over \rho_0}\right)\right)\\
\notag && + \frac{44mA^2}{5\rho_0^2}- \frac{mA^2}{\rho_0^2}\left( \sec^{-1}\left( {r_1\over \rho_0}\right) +\sec^{-1}\left( {r_2\over \rho_0}\right)\right)\,.
\end{eqnarray}
Therefore, as in the previous cases the time delay has the standard value
of general relativity plus the correction term coming from the
parameter A of the scalar field. Note that, in the limit ${m} \rightarrow M_{\odot}$ and $A \rightarrow 0$, the classical general relativity (GR) result is recovered, namely
\begin{equation}
\Delta t_{\text{GR}} = 4M_{\odot} \left[ 1 + \ln\!\left(\frac{4r_1 r_2}{\rho_0^2}\right) \right],
\end{equation}
where $M_{\odot} = 1477~\text{m}$. For a round-trip signal between Earth and Mars, we take 
$r_1 \approx r_2 = 2.25 \times 10^{11}~\text{m}$ as the mean Earth--Mars distance. 
The closest approach to the Sun, $\rho_0$, can be approximated as the sum of the solar radius 
($R_{\odot} \approx 6.960 \times 10^{8}~\text{m}$) and the effective thickness of the solar corona 
($\sim 10^{9}~\text{m}$), yielding $\rho_0 \approx 1.696 \times 10^{9}~\text{m}$. 
Under these assumptions, the resulting time delay is
\begin{equation}
\frac{\Delta t_{\text{GR}}}{c} \approx 240~\mu\text{s}.
\end{equation}

To illustrate the experimental sensitivity, it is worth noting that the uncertainty in the time--delay measurement during the \textit{Viking} mission was only about $10~\text{ns}$~\cite{Straumann}. 
Therefore, if the regular black hole correction term contributes, one obtains
\begin{eqnarray} \notag  
     \Delta t_{\text{R}} &=& \frac{A^2}{\rho_0}
     \Bigg[
        \sec^{-1}\!\left(\frac{r_1}{\rho_0}\right)
        + \sec^{-1}\!\left(\frac{r_2}{\rho_0}\right)
     \Bigg]
     +  \frac{44\,m\,A^2}{5\,\rho_0^2}
     \\[3pt]
     && -\,\frac{m\,A^2}{\rho_0^2}
     \Bigg[
        \sec^{-1}\!\left(\frac{r_1}{\rho_0}\right)
        + \sec^{-1}\!\left(\frac{r_2}{\rho_0}\right)
     \Bigg] .
\end{eqnarray}
For a round--trip signal from Earth to Mars, assuming $r_1 \simeq r_2$, this reduces to
\begin{equation}
\frac{\Delta t_{\text{R}}}{c}
   = \frac{2\,A^2\,(\rho_0 - M_{\odot})}{\rho_0^2\,c}
      \sec^{-1}\!\left(\frac{r_1}{\rho_0}\right)
     + \frac{44\,M_{\odot}\,A^2}{5\,\rho_0^2\,c} .
\end{equation}
Imposing that the correction remains below the experimental uncertainty,
${\Delta t_{\text{R}}}/{c} < 10~\text{ns} = 10^{-8}~\text{s}$, one finds the upper bound
\begin{equation}
A < 40327~\text{m}.
\end{equation}

\subsubsection{Gravitational redshift}

Since the regular black hole is a stationary spacetime there is a time-like Killing vector so that in coordinates adapted to the symmetry the ratio of the measured frequency of a light ray crossing different positions is given by \cite{Kagramanova:2006ax} 
\begin{equation}
{\nu \over \nu_0}=\sqrt{\frac{g_{00}(r)}{g_{00}(r_0)}}\,,
\end{equation}
for  $m/r \ll 1$ and $A/r\ll1$, the above expression yields 
\begin{equation}
{\nu \over \nu_0}\approx  1+ m \left( \frac{1}{r_0}-\frac{1}{r}\right)-\frac{mA^2}{5}\left(\frac{1}{r_0^3}-\frac{1}{r^3}\right)\,.
\end{equation}
Obviously, if we consider the limit ${m} 	\rightarrow M_{\oplus}$, and $A	\rightarrow  0$, we recover the classical result for the Schwarzschild spacetime.
The clock can be compared with an accuracy of $10^{-15}$, the H-maser in the GP-A redshift experiment \cite{Vessot:1980zz} reached an accuracy of $10^{-14}$. Therefore, considering that all observations are well described within Einstein's theory, we conclude that the extra terms of the regular black hole considered must be $\left| \frac{M_{\oplus}A^2}{5}\left(\frac{1}{r_{\oplus}^3}-\frac{1}{r^3}\right)\right| <10^{-14}$. Thus,
\begin{equation}
 A \leq 54819\; \text{m}\,,
\end{equation}
where we assume a clock comparison between Earth and a satellite at 15,000 km height, as in Ref. \cite{Kagramanova:2006ax}.

\section{Lyapunov exponents}
\label{SLE}

Lyapunov exponents are a measure of the average rate at which nearby trajectories converge or diverge in the phase space. Thus, in order to calculate the Lyapunov exponent, we use the Jacobian matrix method \cite{Cardoso:2008bp,Pradhan:2012rkk,Pradhan:2013bli}. So, taking the phase space  $(r,\pi_r)$ the Jacobian matrix $K_{ij}$ is 

\begin{eqnarray}
    K_{11}= \frac{\partial F_1}{\partial r}\,, \quad  K_{12}= \frac{\partial F_1}{\partial \pi_r}\,, \\
    K_{21}= \frac{\partial F_2}{\partial r}\,, \quad K_{22}= \frac{\partial F_2}{\partial \pi_r}\,,
\end{eqnarray}
where $F_1(r,\pi_r)=\frac{dr}{dt}$, and $F_2(r,\pi_r)=\frac{d\pi_r}{dt}$. For circular motion of particles, where $\pi_r=0$, the Jacobian matrix reduces to 
 \begin{equation}
     K_{ij}= \begin{pmatrix}
0 & K_{12}\\
K_{21} & 0 
\end{pmatrix}\,.
 \end{equation}
The eigenvalues of the Jacobian matrix are the Lyapunov exponents $\lambda$
\begin{equation}
    \lambda= \pm \sqrt{K_{12}K_{21}} \,.
\end{equation}
If $\lambda^2>0$ the circular motion is unstable, if $\lambda^2=0$ the circular motion is marginal, and if $\lambda^2<0$ the circular motion is stable. So, for circular null geodesics, the Lyapunov exponent is given by 

\begin{equation}
    \lambda= \pm \sqrt{-\frac{b(r_u)^2}{2E^2} \frac{d^2}{dr^2}V^2(r_u)}\,.\\
\end{equation}

Considering the asymptotically flat regular black hole solution, and evaluating both the potential and the metric function at $r_u$, we obtain:
\begin{equation}
\lambda = \pm \frac{1}{2 A^3} \frac{L}{E} \left( 2A - 3 m \pi + 6 m \arctan \left( \frac{3m}{A}\right) \right)\,.
\end{equation}
 Furthermore, circular null geodesics satisfy the condition $E^2-V^2(r)=0$, which yields $L/E= \pm \omega(r_u)/\sqrt{b(r_u)}$; therefore, the Lyapunov exponent becomes

\begin{equation} \label{LE}
\lambda = \pm \left( \frac{2A- 3m \pi + 6 m \arctan \left( \frac{3m}{A} \right)}{2 A^3}  \right)^{1/2}\,.
\end{equation}
In the following, we drop the $\pm$ sign from the expression above. For small values of the deformation parameter $A$, the Lyapunov exponent behaves as
\begin{equation} \label{lyapunovsmallA}
\lambda \approx \frac{1}{3 \sqrt{3}m}- \frac{A^2}{90 \sqrt{3} m^3} + \frac{79 A^4}{113400 \sqrt{3} m^5} + \mathcal{O}(A^6)\,.
\end{equation}
 For $A \rightarrow 0$ the Lyapunov exponent of the Schwarzschild black hole is recovered, whereas for large values of $A$ the Lyapunov exponent behaves as
\begin{equation}
\lambda \approx \frac{1}{A} - \frac{3 \pi m}{4 A^2} + \mathcal{O}(A^{-3})\,. 
\end{equation}

\begin{figure}[!h]
	\begin{center}
	\includegraphics[width=60mm]{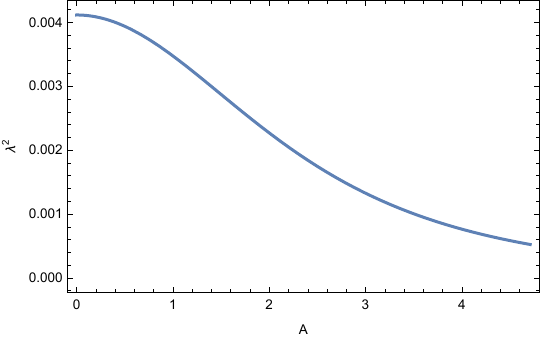}
	\end{center}
	\caption{The Lyapunov exponent $\lambda^2$ as a function of $A$. Here,  $m = 1$.}
	\label{Lyapunov}
\end{figure}

Fig.~\ref{Lyapunov} shows the squared Lyapunov exponent $\lambda^2$ as a function of the parameter $A$ with $m=1$. 
When $A \to 0$, the system approaches the Schwarzschild limit, where the photon sphere exhibits its maximum dynamical instability, recovering the well-known Schwarzschild value $\lambda^2_{\text{Schw}}  \approx  3.72 \times 10^{-2}$. 
For $A \neq 0$, the spacetime corresponds to a regular black hole configuration, free of central singularities. 
In this regime, $\lambda^2$ decreases monotonically with increasing $A$, indicating that the instability of the circular photon orbits weakens as the geometry becomes more regular. 
For instance, at $A = 4.71$, the squared Lyapunov exponent drops to $\lambda^2 \approx  5.21 \times 10^{-4}$, which represents a reduction of more than one order of magnitude compared to the Schwarzschild case. 
This behavior reveals that the parameter $A$ plays a stabilizing role: while the Schwarzschild case ($A=0$) represents the most unstable configuration, larger values of $A$ lead to a progressive stabilization of the photon sphere, with $\lambda^2 \to 0$ marking the onset of marginal stability. 
This trend suggests a deep connection between the regularity of the spacetime and its dynamical stability properties.

On the other hand, although geodesic stability and dynamical stability are conceptually different, our results display a qualitatively consistent trend: in our parametrization the Lyapunov exponent $\lambda^2$ decreases monotonically with increasing $A$, 
indicating that the photon-sphere instability becomes weaker as $A$ grows. This is compatible with the expectation that parameter choices which make the perturbation potential ``more negative'' correspond to stronger instabilities, while moving toward the stable sector identified in Ref.~\cite{Bronnikov:2012ch} softens the instability.

It is worth mentioning that a general upper bound of chaos in quantum systems has been proposed by Maldacena, Shenker and Standford from quantum field theory, which indicates that the Lyapunov exponent $\lambda$, describing the strength of chaos, has a temperature-dependent upper bound \cite{Maldacena:2015waa}

\begin{equation}\label{bound}
    \lambda \leq \frac{2\pi T}{\hbar}.
\end{equation}

This temperature-dependent ansatz was examined by shock wave gedanken experiments \cite{Shenker:2013pqa,Shenker:2013yza} and the AdS/CFT correspondence \cite{Maldacena:1997re}. With the black hole thermodynamic relationship $\kappa = 2\pi T$ and setting $\hbar = 1$, the chaos bound Eq. (\ref{bound}) has an equivalent form at the horizon

\begin{equation}
    \lambda\leq \kappa,
\end{equation}
where $\kappa$ is the surface gravity of black holes.

\section{Connection between Lyapunov exponent and observational signatures}
\label{sec:connection_lyapunov_observables}

An essential aspect of the study of black hole dynamics is the connection between the stability properties of null geodesics and observable quantities. In particular, the Lyapunov exponent, which characterizes the instability timescale of the photon sphere, plays a central role in linking the theoretical framework with measurable astrophysical observables such as the black hole shadow and the deflection angle of light. The Lyapunov exponent $\lambda$ encodes the rate at which nearby null trajectories diverge from the unstable circular orbit at radius $r_{u}$, and thus is directly related to the width of the potential barrier governing photon motion. This connection provides a unified description of various phenomena associated with strong gravitational lensing and the optical appearance of compact objects.

\subsection{Relation with the critical impact parameter} 
\label{sec6}

Having obtained the general expression for the Lyapunov exponent $\lambda$, which quantifies the instability timescale of null circular geodesics, it is instructive to relate this quantity to the critical impact parameter that characterizes the photon sphere. The parameter of impact $\mathcal{B}=L/E$ defines the ratio between the angular momentum and the energy of the photon and determines the boundary that separates the scattering and captured trajectories. For circular null orbits, $\mathcal{B}_u$ takes a specific value that depends solely on the background geometry, $\mathcal{B}_u^2=\omega(r_u)^2/b(r_u)$, where $r_u$ satisfies the photon-sphere condition $b'(r_u)\,\omega(r_u)-2b(r_u)\,\omega'(r_u)=0$. Expressing $\lambda$ in terms of $\mathcal{B}_u$ provides a direct connection between the dynamical instability of the photon orbit and the optical properties of spacetime, which is given by
\begin{eqnarray}
\nonumber \mathcal{B}_u &=& \omega(r_u)/\sqrt{b(r_u)} \\
\nonumber &=& \left( \frac{2 A^3}{2A - 3 \pi m + 6 m \arctan \left( \frac{3m}{A} \right)} \right)^{1/2}  \\
 &=& \frac{1}{|\lambda|}\,.
\end{eqnarray}
This relation is particularly relevant since both $\lambda$ and $\mathcal{B}_u$ can be linked to observable quantities through the frequency and damping timescale of the photon ring \cite{Cardoso:2008bp,Kouniatalis:2025pxs}.

\subsection{Relation with the black hole shadow}
\label{subsec:shadow_connection}

The shadow of a spherically symmetric black hole has been extensively studied in the literature; for a comprehensive review, see \cite{Perlick:2021aok}. The angular radius of the shadow, as perceived by a static observer at a radial coordinate $r_{0}$, is given by
 \begin{equation}
\sin ^2 \alpha_{sh} = \frac{h(r_u)^2}{h(r_0)^2}\,, 
 \end{equation}
where $h(r)$ is given by
\begin{equation}
h(r)^2 = \frac{\omega(r)^2}{b(r)}\,.
\end{equation}
For an observer at a large distance $r_0$, the angular size of the shadow is approximately given by 
\begin{equation}
\alpha_{sh} \approx \frac{h(r_u)}{r_0}\,.
\end{equation}
Evaluating the above expression at the unstable null circular orbit $r_u = 3m$ we obtain: 

\begin{equation} \label{size}
\alpha_{sh} = \frac{1}{r_0} \left( \frac{2 A^3}{2A - 3 \pi m + 6 m \arctan \left( \frac{3m}{A} \right)} \right)^{1/2}\,.
\end{equation}
 For small values of $A$ we obtain: 
\begin{equation}
\alpha_{sh} \approx \frac{3 \sqrt{3}m}{r_0} + \frac{\sqrt{3}A^2}{10 m r_0} + \mathcal{O}(A^4)\,,
\end{equation}
where the first term corresponds to the angular size of the Schwarzschild black hole, and the second term is a correction term which is proportional to the squared deformation parameter.   Therefore, using Eq. (\ref{LE}), we can write Eq. (\ref{size}) as

\begin{equation}
\alpha_{sh} = \frac{1}{r_0 |\lambda|}\,.
\end{equation}
 So, the angular size is inversely proportional to the Lyapunov exponent. The analysis above reveals a direct correspondence between the dynamical instability of the photon sphere and the observable angular size of the black hole shadow. Through Eq.~(\ref{size}), the shadow radius is shown to depend explicitly on the parameter $A$, which governs the deviation from the Schwarzschild geometry and ensures regularity in the center. In particular, as $A$ increases, the Lyapunov exponent $\lambda$ decreases, implying that the photon orbits become less unstable. 
Consequently, since $\alpha_{sh} \propto 1/|\lambda|$, the angular radius of the shadow increases with $A$, leading to a slightly larger apparent size for regular black holes compared to the Schwarzschild case. 
This result establishes a quantitative bridge between the optical appearance and the dynamical properties of spacetime, suggesting that the regularization parameter $A$ not only smooths the central geometry but also stabilizes the photon sphere, thereby producing a more extended and less sharply defined shadow boundary.

\begin{figure}[!h]
\begin{center}
\includegraphics[width=60mm]{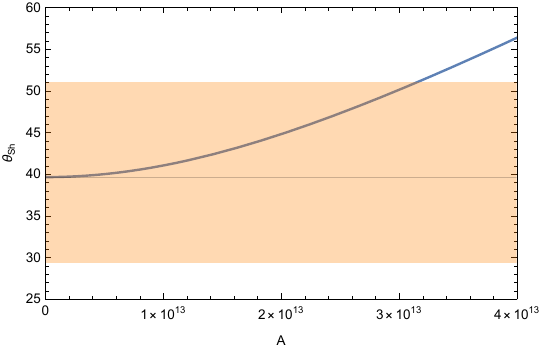}
\includegraphics[width=60mm]{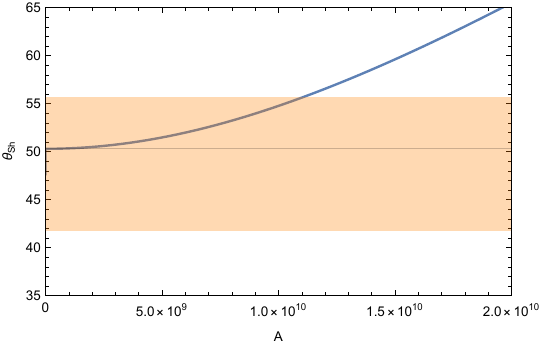}
\end{center}
\caption{Angular diameter $\theta_{sh}$ vs. $A$. The angular diameter increases with $A$. The horizontal line corresponds to the value of the angular diameter for the Schwarzschild black hole. The shaded region denotes the observational bounds. The top panel shows the results for M87*, while the bottom panel shows the corresponding results for Sgr A*}
\label{constrain}
\end{figure}

To constrain the parameter $A$, we employ EHT observations of the black hole shadows of M87* and Sgr A*. The EHT observations have been employed as a probe to distinguish between black hole configurations, see for instance \cite{Vagnozzi:2022moj, Wang:2024lte, Gonzalez:2024ifp}. For M87*, with a mass of $M = 6.5 \times 10^9\, M_\odot $ and a distance of $ r_0 = 16.8\, \mathrm{Mpc} $, the angular diameter $\theta_{sh} = 2 \alpha_{sh}$ is constrained to lie between $ 29.32\, \mu\mathrm{as} $ and $ 51.06\, \mu\mathrm{as} $ \cite{EventHorizonTelescope:2019dse, EventHorizonTelescope:2021dqv}. In the case of Sgr A*, with a mass of $M = 4.0 \times 10^6\, M_\odot $ and a distance of $ r_0 = 8.15\, \mathrm{kpc} $, the measured angular diameter is $ \theta_{sh} = 48.7 \pm 7\, \mu\mathrm{as} $ \cite{EventHorizonTelescope:2022wkp, EventHorizonTelescope:2022xqj}. The corresponding angular diameters are shown in Fig. \ref{constrain}.  These observations yield the constraints $A \lesssim 3.15 \times 10^{13} \, m$ and $A \lesssim 1.1 \times 10^{10} \, m$. Furthermore, the constraint $m \geq 2A/3\pi$ for the existence of the event horizon yields the bounds $A \leq 4.52 \times 10^{13} \, m$ for M87* and $A \leq 2.78 \times 10^{10} \, m$ for Sgr A*.

\section{Conclusions}
\label{FR}

In this work, we have analyzed the geodesic structure and observational properties of asymptotically flat regular black holes sourced by a phantom scalar field with scalar charge $A$. 
This parameter regularizes the central region of the geometry and smoothly deforms the Schwarzschild solution, which is recovered in the limit $A \to 0$. 
The equations of motion for photons were derived and solved to explore the dynamical and optical effects induced by the regularization scale.

The analysis of null trajectories shows that the photon sphere is located at $R_u=\omega(3m)$, while the invariant areal radius of the horizon $\omega(r_+)$ increases with increasing $A$, indicating that the causal region becomes less compact. 
The effective potential and coordinate-time profiles demonstrate that the gravitational time dilation and the trapping efficiency are progressively attenuated for larger values of $A$.

From the confrontation with classical Solar System tests, we have obtained quantitative bounds on the parameter $A$. 
The analysis of the light deflection yields $A = 4.63\times10^{4}\,\mathrm{m}$ for the prograde case and $A = 4.41\times10^{4}\,\mathrm{m}$ for the retrograde one. 
The radar echo delay provides a slightly stronger constraint, $A = 4.03\times10^{4}\,\mathrm{m}$, while the gravitational redshift test gives $A \leq 5.48\times10^{4}\,\mathrm{m}$. 
These independent limits confirm that the scalar charge must be extremely small in the weak-field regime, ensuring full compatibility with general relativity within the current experimental precision. On the other hand, Solar-System observations constrain $A$ in the weak-field regime, while the stability analysis of Ref.~\cite{Bronnikov:2012ch} selects the stable 
solution branch through the critical ratio $A/m=3\pi/2$. The ranges obtained from Solar-System tests include this stability value, demonstrating their compatibility.

In the strong-field domain, the Lyapunov analysis of photon orbits shows that the instability of the circular null geodesic weakens as $A$ increases, linking the regularization of spacetime to a gradual stabilization of photon trajectories. 
We have established precise relations between the Lyapunov exponent and optical observables, namely the critical impact parameter $B_u = 1/|\lambda|$ and the angular size of the shadow $\alpha_{\mathrm{sh}} = 1/(r_{0}|\lambda|)$. 
These correspondences indicate that more regular configurations produce slightly larger and smoother black-hole shadows compared to the Schwarzschild limit. 
In addition, constraints derived from Event Horizon Telescope observations further bound the scalar parameter to $A \lesssim 3.15\times10^{13}\,\mathrm{m}$ for M87* and $A \lesssim 1.1\times10^{10}\,\mathrm{m}$ for Sgr~A*, reinforcing the consistency of the model with current strong-field observations.

In summary, the scalar parameter $A$ acts as a regulator, controlling both the causal and optical properties of spacetime. 
Although its value must remain small to satisfy Solar System constraints, its effects become relevant in the strong-field regime, where it influences the Lyapunov instability, photon capture, and shadow size. 
These results provide a coherent framework to interpret future high-resolution observations from the Event Horizon Telescope and related interferometric projects aimed at probing the near-horizon geometry of regular black holes.

\begin{acknowledgments}

We thank the anonymous referee for valuable comments and suggestions. Y. V. acknowledges support by the Direcci\'on de Investigaci\'on y Desarrollo de la Universidad de La Serena, Grant No. PR25538511.

\end{acknowledgments}


\begin{thebibliography}{99}

\bibitem{steinhardt}
A.~Upadhye, M.~Ishak and P.~J.~Steinhardt,
``Dynamical dark energy: Current constraints and forecasts,''
Phys. Rev. D \textbf{72} (2005), 063501
[arXiv:astro-ph/0411803 [astro-ph]].


\bibitem{tegmark}
Y.~Wang and M.~Tegmark,
``New dark energy constraints from supernovae, microwave background and galaxy clustering,''
Phys. Rev. Lett. \textbf{92} (2004), 241302
[arXiv:astro-ph/0403292 [astro-ph]].

\bibitem{seljak}
U.~Seljak \textit{et al.} [SDSS],
``Cosmological parameter analysis including SDSS Ly-alpha forest and galaxy bias: Constraints on the primordial spectrum of fluctuations, neutrino mass, and dark energy,''
Phys. Rev. D \textbf{71} (2005), 103515
[arXiv:astro-ph/0407372 [astro-ph]].

\bibitem{hannestad}
S.~Hannestad and E.~Mortsell,
``Cosmological constraints on the dark energy equation of state and its evolution,''
JCAP \textbf{09} (2004), 001
[arXiv:astro-ph/0407259 [astro-ph]].

\bibitem{star03}
U.~Alam, V.~Sahni, T.~D.~Saini and A.~A.~Starobinsky,
``Is there supernova evidence for dark energy metamorphosis ?,''
Mon. Not. Roy. Astron. Soc. \textbf{354} (2004), 275
[arXiv:astro-ph/0311364 [astro-ph]].


\bibitem{chandra}
S.~W.~Allen, R.~W.~Schmidt, H.~Ebeling, A.~C.~Fabian and L.~van Speybroeck,
``Constraints on dark energy from Chandra observations of the largest relaxed galaxy clusters,''
Mon. Not. Roy. Astron. Soc. \textbf{353} (2004), 457
[arXiv:astro-ph/0405340 [astro-ph]].

\bibitem{sen}
A. Sen,
``Rolling tachyon,''
JHEP \textbf{04} (2002), 048
[arXiv:hep-th/0203211 [hep-th]].
``Tachyon matter,''
JHEP \textbf{07} (2002), 065
[arXiv:hep-th/0203265 [hep-th]].

\bibitem{gorini}
V.~Gorini, A.~Y.~Kamenshchik, U.~Moschella and V.~Pasquier,
``Tachyons, scalar fields and cosmology,''
Phys. Rev. D \textbf{69} (2004), 123512
[arXiv:hep-th/0311111 [hep-th]].

\bibitem{fara05}
V.~Faraoni,
``Phantom cosmology with general potentials,''
Class. Quant. Grav. \textbf{22} (2005), 3235-3246
[arXiv:gr-qc/0506095 [gr-qc]].

\bibitem{no03}
S.~Nojiri and S.~D.~Odintsov,
``Quantum de Sitter cosmology and phantom matter,''
Phys. Lett. B \textbf{562} (2003), 147-152
[arXiv:hep-th/0303117 [hep-th]].

\bibitem{trod}
S.~M.~Carroll, M.~Hoffman and M.~Trodden,
``Can the dark energy equation-of-state parameter $w$ be less than $-1$?,''
Phys. Rev. D \textbf{68} (2003), 023509
[arXiv:astro-ph/0301273 [astro-ph]].



\bibitem{Bronnikov:2005gm}
K.~A.~Bronnikov and J.~C.~Fabris,
``Regular phantom black holes,''
Phys. Rev. Lett. \textbf{96} (2006), 251101
[arXiv:gr-qc/0511109 [gr-qc]].
K.~A.~Bronnikov,
``Scalar fields as sources for wormholes and regular black holes,''
Particles \textbf{1} (2018) no.1, 56-81
[arXiv:1802.00098 [gr-qc]].



\bibitem{dym92}
I.~Dymnikova,
``Vacuum nonsingular black hole,''
Gen. Rel. Grav. \textbf{24}, 235-242 (1992).

\bibitem{ned01}
K.~A.~Bronnikov,
``Regular magnetic black holes and monopoles from nonlinear electrodynamics,''
Phys. Rev. D \textbf{63} (2001), 044005
[arXiv:gr-qc/0006014 [gr-qc]].

\bibitem{bdd03}
K.~A.~Bronnikov, A.~Dobosz and I.~G.~Dymnikova,
``Nonsingular vacuum cosmologies with a variable cosmological term,''
Class. Quant. Grav. \textbf{20} (2003), 3797-3814
[arXiv:gr-qc/0302029 [gr-qc]].


\bibitem{Barrientos:2025rjn}
J.~Barrientos, A.~Cisterna, M.~Hassaine, K.~M{\"u}ller and K.~Pallikaris,
``A new exact rotating spacetime in vacuum: The Kerr{\textendash}Levi-Civita spacetime,''
Phys. Lett. B \textbf{871} (2025), 140035
[arXiv:2506.07166 [gr-qc]].



\bibitem{Karakasis:2023hni}
T.~Karakasis, N.~E.~Mavromatos and E.~Papantonopoulos,
``Regular compact objects with scalar hair,''
Phys. Rev. D \textbf{108}, no.2, 024001 (2023)
[arXiv:2305.00058 [gr-qc]].


\bibitem{Farrah:2023opk}
D.~Farrah, K.~S.~Croker, G.~Tarl\'e, V.~Faraoni, S.~Petty, J.~Afonso, N.~Fernandez, K.~A.~Nishimura, C.~Pearson and L.~Wang, \textit{et al.}
``Observational Evidence for Cosmological Coupling of Black Holes and its Implications for an Astrophysical Source of Dark Energy,''
Astrophys. J. Lett. \textbf{944}, no.2, L31 (2023)
[arXiv:2302.07878 [astro-ph.CO]].

\bibitem[{Akiyama {et~al.}
(2022)Akiyama, Alberdi, Alef, Algaba, Anantua, Asada,
  Azulay, Bach, Baczko, Ball, {et~al.}}]{akiyama2022first}
Akiyama, K., Alberdi, A., Alef, W., {et~al.} 2022, The Astrophysical Journal
  Letters, 930, L12.


  
\bibitem{Gonzalez:2015jna}
P.~A.~Gonzalez, M.~Olivares and Y.~Vasquez,
``Motion of particles on a Four-Dimensional Asymptotically AdS Black Hole with Scalar Hair,''
Eur. Phys. J. C \textbf{75} (2015) no.10, 464
[arXiv:1507.03610 [gr-qc]].

\bibitem{Ramos:2021jta}
A.~Ramos, C.~Arias, R.~Avalos and E.~Contreras,
``Geodesic motion around hairy black holes,''
Annals Phys. \textbf{431} (2021), 168557
[arXiv:2107.01146 [gr-qc]].

\bibitem{Heydari-Fard:2021pjc}
M.~Heydari-Fard, M.~Heydari-Fard and H.~R.~Sepangi,
``Null geodesics and shadow of hairy black holes in Einstein-Maxwell-dilaton gravity,''
Phys. Rev. D \textbf{105} (2022) no.12, 124009
[arXiv:2110.02713 [gr-qc]].



\bibitem{Theodosopoulos:2023ice}
D.~P.~Theodosopoulos, T.~Karakasis, G.~Koutsoumbas and E.~Papantonopoulos,
``Motion of particles around a magnetically charged Euler{\textendash}Heisenberg black hole with scalar hair and the Event Horizon Telescope,''
Eur. Phys. J. C \textbf{84} (2024) no.6, 592
[arXiv:2311.02740 [gr-qc]].


\bibitem{Chen:2024luw}
H.~Chen, W.~Fan and X.~Y.~Chew,
``Geodesic motion of test particles around the scalar hairy black holes with asymmetric vacua,''
Eur. Phys. J. C \textbf{85} (2025) no.3, 338
[arXiv:2411.00565 [gr-qc]].


\bibitem{Ovalle:2020kpd}
J.~Ovalle, R.~Casadio, E.~Contreras and A.~Sotomayor,
``Hairy black holes by gravitational decoupling,''
Phys. Dark Univ. \textbf{31}, 100744 (2021)
[arXiv:2006.06735 [gr-qc]].


\bibitem{Nekouee:2025zvp}
Z.~Nekouee, S.~K.~Narasimhamurthy, B.~R.~Yashwanth and T.~Sanjay,
``Exploring null geodesic of Finslerian hairy black hole,''
Class. Quant. Grav. \textbf{42} (2025) no.4, 045002


\bibitem{Carvajal:2025ucx}
D.~A.~Carvajal, P.~A.~Gonz{\'a}lez, M.~Olivares, E.~Papantonopoulos and Y.~V{\'a}squez,
``Massive Particle Motion Around Horndeski Black Holes,''
[arXiv:2507.09819 [gr-qc]].

\bibitem{Carvajal:2025emj}
D.~A.~Carvajal, P.~A.~Gonz{\'a}lez, M.~Olivares, E.~Papantonopoulos and Y.~V{\'a}squez,
``Study of null geodesics and their stability in Horndeski black holes,''
Eur. Phys. J. C \textbf{85} (2025) no.9, 978
[arXiv:2503.02083 [gr-qc]].


\bibitem{Gonzalez:2020vzl}
P.~A.~Gonz{\'a}lez, M.~Olivares, E.~Papantonopoulos and Y.~V{\'a}squez,
``Constraints on scalar{\textendash}tensor theory of gravity by solar system tests,''
Eur. Phys. J. C \textbf{80} (2020) no.10, 981
[arXiv:2002.03394 [gr-qc]].




\bibitem{Abbas:2014oua}
G.~Abbas and U.~Sabiullah,
``Geodesic Study of Regular Hayward Black Hole,''
Astrophys. Space Sci. \textbf{352} (2014), 769-774
[arXiv:1406.0840 [gr-qc]].


\bibitem{Stuchlik:2014qja}
Z.~Stuchl{\'\i}k and J.~Schee,
``Circular geodesic of Bardeen and Ayon{\textendash}Beato{\textendash}Garcia regular black-hole and no-horizon spacetimes,''
Int. J. Mod. Phys. D \textbf{24} (2014) no.02, 1550020
[arXiv:1501.00015 [astro-ph.HE]].


\bibitem{Azam:2017adt}
M.~Azam, G.~Abbas, S.~Sumera and A.~R.~Nizami,
``Geodesic structure of magnetically charged regular black hole,''
Int. J. Geom. Meth. Mod. Phys. \textbf{14} (2017) no.09, 1750120

\bibitem{Azam:2017izk}
M.~Azam, G.~Abbas and S.~Sumera,
``Geodesic motion around regular magnetic black hole in non-minimal Einstein{\textendash}Yang{\textendash}Mills theory,''
Can. J. Phys. \textbf{95} (2017) no.11, 1062-1067



\bibitem{Becerril:2020fek}
R.~Becerril, S.~Valdez-Alvarado, U.~Nucamendi, P.~Sheoran and J.~M.~D{\'a}vila,
``Mass parameter and the bounds on redshifts and blueshifts of photons emitted from geodesic particle orbiting in the vicinity of regular black holes,''
Phys. Rev. D \textbf{103} (2021) no.8, 084054
[arXiv:2007.15300 [gr-qc]].



\bibitem{Zhou:2022yio}
T.~Zhou and L.~Modesto,
``Geodesic incompleteness of some popular regular black holes,''
Phys. Rev. D \textbf{107} (2023) no.4, 044016
[arXiv:2208.02557 [gr-qc]].


\bibitem{Bautista-Olvera:2019blb}
B.~Bautista-Olvera, J.~C.~Degollado and G.~German,
``Geodesic structure of a rotating regular black hole,''
Gen. Rel. Grav. \textbf{55} (2023) no.5, 66
[arXiv:1908.01886 [gr-qc]].


\bibitem{Xi:2023oib}
Z.~Xi, C.~Wu and W.~Guo,
``Geodesic structure of test particles in different regular black holes,''
Can. J. Phys. \textbf{102} (2024) no.12, 615-622
[arXiv:2309.12932 [gr-qc]].


\bibitem{primary}  D.~C.~Zou and Y.~S.~Myung,
``Black hole with primary scalar hair in Einstein-Weyl-Maxwell-conformal scalar theory,''
Phys. Rev. D \textbf{101}, no.8, 084021 (2020)
[arXiv:2001.01351 [gr-qc]];

A.~Anabalon, F.~Canfora, A.~Giacomini and J.~Oliva,
``Black Holes with Primary Hair in gauged N=8 Supergravity,''
JHEP \textbf{06}, 010 (2012)
[arXiv:1203.6627 [hep-th]];

S.~Mignemi and D.~L.~Wiltshire,
``Multi-scalar black holes with contingent primary hair: Mechanics and stability,''
Phys. Rev. D \textbf{70}, 124012 (2004)
[arXiv:hep-th/0408215 [hep-th]];


K.~Dialektopoulos, T.~Papanikolaou and V.~Zarikas,
``Primordial black holes as cosmic expansion accelerators,''
Phys. Lett. B \textbf{870}, 139948 (2025)
[arXiv:2502.18352 [gr-qc]];

S.~Mignemi,
``Primary scalar hair in dilatonic theories with modulus fields,''
Phys. Rev. D \textbf{62}, 024014 (2000)
[arXiv:gr-qc/9910041 [gr-qc]];

P.~A.~Gonz\'alez, E.~Papantonopoulos, J.~Saavedra and Y.~V\'asquez,
``Four-Dimensional Asymptotically AdS Black Holes with Scalar Hair,''
JHEP \textbf{12} (2013), 021
[arXiv:1309.2161 [gr-qc]].

\bibitem{Bronnikov:2012ch}
K.~A.~Bronnikov, R.~A.~Konoplya and A.~Zhidenko,
``Instabilities of wormholes and regular black holes supported by a phantom scalar field,''
Phys. Rev. D \textbf{86} (2012), 024028
[arXiv:1205.2224 [gr-qc]].

\bibitem{Ding:2013vta}
C.~Ding, C.~Liu, Y.~Xiao, L.~Jiang and R.~G.~Cai,
``Strong gravitational lensing in a black-hole spacetime dominated by dark energy,''
Phys. Rev. D \textbf{88} (2013) no.10, 104007
[arXiv:1308.5035 [gr-qc]].

\bibitem{Ditta2020}
A. Ditta and G. Abbas, “Circular orbits and accretion process near a regular phantom black hole,” Gen. Rel. Grav. 52 (2020) no.8, 77.




  \bibitem{Chandrasekhar:579245}
S. Chandrasekhar,
``The mathematical theory of black holes", Oxford University Press, 2002.



\bibitem{Cruz:2004ts}
N.~Cruz, M.~Olivares and J.~R.~Villanueva,
``The Geodesic structure of the Schwarzschild anti-de Sitter black hole,''
Class. Quant. Grav. \textbf{22} (2005), 1167-1190
[arXiv:gr-qc/0408016 [gr-qc]].


  
 
\bibitem{Villanueva:2018kem}
  J.~R.~Villanueva, F.~Tapia, M.~Molina and M.~Olivares,
  ``Null paths on a toroidal topological black hole in conformal Weyl gravity,''
  Eur.\ Phys.\ J.\ C {\bf 78} (2018) no.10,  853
  [arXiv:1808.04298 [gr-qc]].

\bibitem{wald}
Wald R.M.:
General relativity.
The University Chicago Press, Chicago (1984).


\bibitem{Straumann}
N. Straumann,
``General relativity and relativistic astrophysics'',
 Springer, Berlin, 1984.



\bibitem{Roy:2019ijp}
S.~Roy and A.~K.~Sen,
``Study of gravitational deflection of light ray,''
J. Phys. Conf. Ser. \textbf{1330} (2019) no.1, 012002.

\bibitem{Fathi:2025byw}
M.~Fathi and A.~{\"O}vg{\"u}n,
``Black hole with global monopole charge in self-interacting Kalb-Ramond field,''
Eur. Phys. J. Plus \textbf{140} (2025) no.4, 280
[arXiv:2501.09899 [gr-qc]].

\bibitem{Kagramanova:2006ax}
  V.~Kagramanova, J.~Kunz and C.~Lammerzahl,
  ``Solar system effects in Schwarzschild-de Sitter spacetime,''
  Phys.\ Lett.\ B {\bf 634}, 465 (2006)
  [gr-qc/0602002].


\bibitem{Vessot:1980zz}
  R.~F.~C.~Vessot {\it et al.},
  ``Test of Relativistic Gravitation with a Space-Borne Hydrogen Maser,''
  Phys.\ Rev.\ Lett.\  {\bf 45}, 2081 (1980).



  
   \bibitem{Cardoso:2008bp}
   V.~Cardoso, A.~S.~Miranda, E.~Berti, H.~Witek and V.~T.~Zanchin,
   Geodesic stability, Lyapunov exponents and quasinormal modes,
   Phys. Rev. D \textbf{79} (2009) no.6, 064016.
   [arXiv:0812.1806 [hep-th]].

\bibitem{Pradhan:2012rkk}
P.~Pradhan,
``Stability analysis and quasinormal modes of Reissner\textendash{}Nordstr\o{}m space-time via Lyapunov exponent,''
Pramana \textbf{87} (2016) no.1, 5
[arXiv:1205.5656 [gr-qc]].

\bibitem{Pradhan:2013bli}
P.~P.~Pradhan,
``Lyapunov Exponent and Charged Myers Perry Spacetimes,''
Eur. Phys. J. C \textbf{73} (2013) no.6, 2477
[arXiv:1302.2536 [gr-qc]].

\bibitem{Maldacena:2015waa}
J.~Maldacena, S.~H.~Shenker and D.~Stanford,
``A bound on chaos,''
JHEP \textbf{08} (2016), 106
[arXiv:1503.01409 [hep-th]].

\bibitem{Shenker:2013pqa}
S.~H.~Shenker and D.~Stanford,
``Black holes and the butterfly effect,''
JHEP \textbf{03} (2014), 067
[arXiv:1306.0622 [hep-th]].


\bibitem{Shenker:2013yza}
S.~H.~Shenker and D.~Stanford,
``Multiple Shocks,''
JHEP \textbf{12} (2014), 046
[arXiv:1312.3296 [hep-th]].


\bibitem{Maldacena:1997re}
J.~M.~Maldacena,
``The Large N limit of superconformal field theories and supergravity,''
Adv. Theor. Math. Phys. \textbf{2} (1998), 231-252
[arXiv:hep-th/9711200 [hep-th]].




\bibitem{Kouniatalis:2025pxs}
G.~Kouniatalis, P.~A.~Gonz{\'a}lez, E.~Papantonopoulos and Y.~V{\'a}squez,
``Near-horizon Instabilities and Anomalous Decay Rate of Quasinormal Modes in Weyl Black Holes,''
[arXiv:2509.05164 [gr-qc]].




\bibitem{Perlick:2021aok}
V.~Perlick and O.~Y.~Tsupko,
``Calculating black hole shadows: Review of analytical studies,''
Phys. Rept. \textbf{947}, 1-39 (2022)
[arXiv:2105.07101 [gr-qc]].

\bibitem{Vagnozzi:2022moj}
S.~Vagnozzi, R.~Roy, Y.~D.~Tsai, L.~Visinelli, M.~Afrin, A.~Allahyari, P.~Bambhaniya, D.~Dey, S.~G.~Ghosh and P.~S.~Joshi, \textit{et al.}
``Horizon-scale tests of gravity theories and fundamental physics from the Event Horizon Telescope image of Sagittarius A,''
Class. Quant. Grav. \textbf{40}, no.16, 165007 (2023)
[arXiv:2205.07787 [gr-qc]].


 \bibitem{Wang:2024lte}
X.~J.~Wang, Y.~Meng, X.~M.~Kuang and K.~Liao,
``Distinguishing black holes with and without spontaneous scalarization in Einstein-scalar-Gauss{\textendash}Bonnet theories via optical features,''
Eur. Phys. J. C \textbf{84}, no.12, 1243 (2024)
[arXiv:2409.20200 [gr-qc]].


\bibitem{Gonzalez:2024ifp}
P.~A.~Gonz{\'a}lez, E.~Papantonopoulos, J.~Robledo and Y.~V{\'a}squez,
``Nonlinear scalarization of Schwarzschild black holes in scalar-torsion teleparallel gravity,''
Phys. Rev. D \textbf{111}, no.4, 044064 (2025)
[arXiv:2407.13557 [gr-qc]].

\bibitem{EventHorizonTelescope:2019dse}
K.~Akiyama \textit{et al.} [Event Horizon Telescope],
``First M87 Event Horizon Telescope Results. I. The Shadow of the Supermassive Black Hole,''
Astrophys. J. Lett. \textbf{875}, L1 (2019)
[arXiv:1906.11238 [astro-ph.GA]].

\bibitem{EventHorizonTelescope:2021dqv}
P.~Kocherlakota \textit{et al.} [Event Horizon Telescope],
``Constraints on black-hole charges with the 2017 EHT observations of M87*,''
Phys. Rev. D \textbf{103}, no.10, 104047 (2021)
[arXiv:2105.09343 [gr-qc]].


\bibitem{EventHorizonTelescope:2022wkp}
K.~Akiyama \textit{et al.} [Event Horizon Telescope],
``First Sagittarius A* Event Horizon Telescope Results. I. The Shadow of the Supermassive Black Hole in the Center of the Milky Way,''
Astrophys. J. Lett. \textbf{930}, no.2, L12 (2022)
[arXiv:2311.08680 [astro-ph.HE]].


\bibitem{EventHorizonTelescope:2022xqj}
K.~Akiyama \textit{et al.} [Event Horizon Telescope],
``First Sagittarius A* Event Horizon Telescope Results. VI. Testing the Black Hole Metric,''
Astrophys. J. Lett. \textbf{930}, no.2, L17 (2022)
[arXiv:2311.09484 [astro-ph.HE]].





\end{thebibliography}
\end{document}